\documentclass[twocolumn,pra,superscriptaddress,amsmath,amssymb,mathrsfs,showpacs,floatfix]{revtex4-1}
\usepackage{tabularx}
\usepackage{graphicx}
\usepackage{braket}
\usepackage[utf8]{inputenc}
\usepackage[colorlinks=true,citecolor=blue,linkcolor=blue]{hyperref}
\usepackage{dcolumn}
\usepackage{bm}
\usepackage{epsfig}
\usepackage{rotating}
\usepackage{color}

\begin{document}

\title{Quantum control of ro-vibrational dynamics and application to light-induced molecular chirality}

\author{Monika Leibscher}
\affiliation{Dahlem Center for Complex Quantum Systems and Fachbereich Physik, Freie Universit\"{a}t Berlin,
	Arnimallee 14, 14195 Berlin, Germany}

\author{Eugenio Pozzoli}
\affiliation{Univ Rennes, CNRS, IRMAR - UMR 6625, F-35000 Rennes, France}
\author{Alexander Blech}
\affiliation{Dahlem Center for Complex Quantum Systems and Fachbereich Physik, Freie Universit\"{a}t Berlin,
	Arnimallee 14, 14195 Berlin, Germany}

\author{Mario Sigalotti} 
\affiliation{
	Laboratoire Jacques-Louis Lions, Sorbonne Universit\'e, Universit\'e de Paris, CNRS, Inria, Paris, France}

\author{Ugo Boscain} 
\affiliation{
	Laboratoire Jacques-Louis Lions, Sorbonne Universit\'e, Universit\'e de Paris, CNRS, Inria, Paris, France}

\author{Christiane P. Koch} 
\thanks{Corresponding author.}
\email{christiane.koch@fu-berlin.de}
\affiliation{Dahlem Center for Complex Quantum Systems and Fachbereich Physik, Freie Universit\"{a}t Berlin,
	Arnimallee 14, 14195 Berlin, Germany}

\date{\today} 
 
 \begin{abstract}
  Achiral molecules can be made temporarily chiral 
  by excitation with electric fields, in the sense that an average over molecular orientations displays a net chiral signal [Tikhonov et al., Sci. Adv. 8, eade0311 (2022)]. Here, we go beyond the assumption of molecular orientations to remain fixed during the excitation process. Treating both rotations and vibrations quantum mechanically, we identify conditions for the creation of chiral vibrational wavepackets -- with net chiral signals -- in ensembles of achiral molecules which are initially randomly oriented. Based on the analysis of symmetry and controllability, we derive excitation schemes for the creation of chiral wavepackets using a combination of (a)  microwave and IR pulses and (b) a static field and a sequence of IR pulses. These protocols leverage quantum rotational dynamics for pump-probe spectroscopy of chiral vibrational dynamics, extending the latter to regions of the electromagnetic spectrum other than the UV. 
\end{abstract}

\maketitle

\section{Introduction}
The insight that molecular chirality may be explored in gas phase ensembles of molecules with random orientations~\cite{LuxAngewandte12,PattersonNature13} has triggered a surge of experimental activity, and chiral molecules interacting with light in the electric dipole approximation have become a central focus of current AMO research, both experimental~\cite{CireasaNatPhys15,ShubertAngewandte14,LobsigerJPCL15,MilnerPRL2019,Lee21,Singh_Angw_2023,FaccialaPRX23} and theoretical~\cite{GoetzJCP17,Demekhin_PRL_2018,TutunnikovJPCL18,LehmannJCP18,Leibscher19,OrdonezPRA19,GoetzPRL19,NeufeldPRX19,VogwellSciAdv21}. 
Methods like photoelectron dichroism \cite{LuxAngewandte12,CireasaNatPhys15,Kastner_JCP_2017,Beaulieu_Nature_2018,Ranecky_PCCP_2022,Comby_PCCP_2023,FaccialaPRX23}, chiral-sensitive high-harmonic generation~\cite{BakyushevaPRX18}, laser induced enantiomer selective molecular orientation~\cite{TutunnikovJPCL18,MilnerPRL2019}  or microwave three-wave mixing~\cite{PattersonNature13,ShubertAngewandte14,LobsigerJPCL15,Domingos2020,Lee21,Singh_Angw_2023} allow to discriminate between enantiomers of chiral molecules in the gas phase. At the same time, it is not yet clear what ultimately determines the magnitude of these chiral signatures. One way to approach this question is to imprint chirality onto achiral molecules~\cite{OwensPRL16,Tikhonov22} or atoms~\cite{Ilchen_PRL_2017,Ordonez_PRA_2019_1,GrumG_PRA_2019,Buhmann_2021,MayerPRL22}.

For example, an achiral molecule can become temporarily chiral if the nuclei are distorted from their achiral equilibrium configuration by exciting nuclear vibrations and the oscillation between chiral and achiral structures can be measured by photoelectron circular dichroism \cite{Tikhonov22}. 
When starting from a planar molecule, Raman excitation of an out-of-plane normal mode in the presence of a static electric field has been proposed to create a chiral vibrational wavepacket \cite{Tikhonov22}. In general, the interaction of three orthogonal components of the molecular (transition) dipole moment with electric fields with three orthogonal polarization directions is sufficient to induce chirality in an achiral structure and yield a net chiral signal when averaged over random orientations  \cite{Tikhonov22}. This is in full analogy to the conditions for enantiomer-selective response in a sample of randomly oriented chiral molecules with light-matter interaction in the electric dipole approximation \cite{OrdonezPRA18}. 
In both cases~\cite{Tikhonov22,OrdonezPRA18}, the conditions for enantiomer-sensitivity have been derived under the assumption that molecular rotations are frozen during the interaction and can be described by a classical probability distribution over Euler angles. For ultrafast Raman excitation, this is a valid assumption but rotations and vibrations can also be driven by much slower processes, e.g. by long, narrow-band IR pulses. In the latter case, the molecules rotate while external fields excite molecular vibrations, and the rotation affects even purely vibrational observables. On the one hand, the rotation may result in decoherence of the vibrational superpositions. On the other hand, it may allow for new excitation processes to create chiral vibrational wavepackets.
Here, we identify and describe these new routes to imprint temporal chirality onto achiral molecules. To this end, we present a full quantum mechanical treatment of the ro-vibrational dynamics which allows us to generalize the conditions for creating temporal chiral structures to rotating molecules and to distinguish between classical and quantum mechanical routes to induce chirality. 

In order to determine the conditions for creating chirality in randomly rotating molecules, we apply two methods:  (i) We employ the symmetry properties of asymmetric top rotors \cite{Bunker}, extending an earlier symmetry analysis for rigid chiral molecules \cite{Leibscher19} to ro-vibrational dynamics. (ii) We analyze the controllability of a vibrating quantum rotor. Controllability analysis answers the question whether it is possible or not to reach a control target with a given set of external fields \cite{Alessandro2008}.
Analyzing the controllability of rotational systems is challenging due to the inherent degeneracies. In the rigid rotor limit where vibrations are ignored, controllability properties have been derived for linear rotors \cite{Judson1990,BCS,Chambrion23} as well as symmetric \cite{Boscain21,Chambrion22} and asymmetric tops \cite{Leibscher22,Pozzoli21,Pozzoli22}. 
Recently developed graphical methods to analyze the controllability  \cite{BCCS,CMSB,graphs} have been proven helpful to analyze controllability of quantum rotors \cite{Pozzoli21,Leibscher22}. Here, we apply these methods to vibrating rotors. 
This analysis allows us to identify the new excitation processes for the creation of chiral vibrational wavepackets which we verify by numerical simulations of the ro-vibrational dynamics. 

The paper is organized as follows:
 In Section \ref{sec:model}, we present the theoretical framework for describing vibrational observables in a driven ro-vibrational quantum system. General conditions for exciting coherent vibrational wavepackets in randomly oriented molecules based on the symmetry as well as the formulation of the control problem are discussed in Section \ref{sec:condition}. In Sections \ref{sec:example1} and \ref{sec:example2} we present two different strategies to create a chiral vibrational wavepacket. The first scheme combines a purely rotational excitation using microwave pulses with an IR pulse that induces ro-vibrational transitions. This scheme is described in Section \ref{sec:example1}. In Section \ref{sec:example2} we demonstrate that a chiral wavepacket can also be excited with three IR pulses in combination with a static electric field. 
In Section~\ref{sec:conclusions} we summarize our findings and conclude.

\section{Vibrational wavepackets in rotating molecules: Theoretical framework}\label{sec:model}

In order to describe vibrational excitation in randomly oriented molecules, we model the molecular Hamiltonian as 
\begin{equation}
H_{0}= \sum_{\nu} \sum_j \left ( E_\nu^{vib} + E_j^{rot}\right )
\ket{\nu} \ket{\phi_j}\bra{\phi_j}\bra{\nu},
\end{equation}
where we neglect any ro-vibrational coupling. We consider a single vibrational mode, represented by the operator $\hat{\chi}$, and the vibrational eigenstates $\ket{\nu}$ along this normal. The eigenstates of a rigid  top  are denoted by $\ket{\phi_j}$, where $j$ indicates the quantum numbers of the rigid rotor. 
The vibrational energies are denoted by
$E_\nu^{vib}$, and $E_j^{rot}$ are the eigenvalues of the rigid rotor.
The molecules evolve according to the time-dependent Schr\"odinger equation,
\begin{equation}
i \hbar \frac{\partial}{\partial t} \ket{\psi(t)} = \left ( H_0 + H_{int}(t) \right ) \ket{\psi(t)}.
\label{eq:TDSE}
\end{equation}
The interaction between the molecules and a set of electric fields
\begin{equation}
{\bf E}_i=\bm{e}_i {\cal E}_i u_i(t)
\label{eq:electric_fields}
\end{equation}
is described in the electric dipole approximation as
\begin{equation}
H_{int}(t) = - \sum_i  u_i(t) H_i 
\label{eq:Hint_t}
\end{equation}
with
\begin{equation}
H_i = {\cal E}_i \bm{\mu} \cdot {\bm R}(\gamma_R) \cdot \bm{e}_i.
\label{eq:Hint}
\end{equation}
{\color{black} Here $\bm{e}_i$ is either $\bm{e}_x,\bm{e}_y$ or $\bm{e}_z$ and denotes the polarization of the $i$th electric field  in the space-fixed coordinate system. The maximal field strength is given by ${\cal E}_i$, and $u_i(t) = s_i(t) \cos(\omega_i t + \phi_i)$ is the time-dependence of the electric field with the dimensionless envelope $s_i(t)$,  frequency $\omega_i$ and phase $\phi_i$.} The molecular dipole moment ${\bm \mu}=(\mu_a,\mu_b,\mu_c)^T$ is given in molecule-fixed coordinates, and the rotation matrix ${\bm R}(\gamma_R)$ transforms between the space-fixed and molecule-fixed coordinate system. It depends on the Euler angles $\gamma_R=(\theta,\psi,\varphi)$.
The components of the dipole moment, $\mu_\alpha$, $\alpha=a,b,c$, are functions of the nuclear coordinates, i.e, the normal mode $\chi$. The interaction with the electric field thus couples molecular vibrations with the rotational degrees of freedom. 

Upon excitation with electric fields, a ro-vibrational wavepacket of the form
\begin{equation}
\ket{\psi(t)} =\sum_j \sum_\nu c_{\nu,j} (t) \ket{\nu} \ket{\phi_j}
\end{equation}
is excited. We now consider an operator that acts only on the vibrational subspace, e.g. the coordinate ${\hat \chi}$ of a normal mode of the molecule. The expectation value of ${\hat \chi}$, which includes integrating over the rotational degrees of freedom, is then given by
\begin{equation}
\langle {\hat \chi} \rangle (t) = \bra{\psi(t)} {\hat \chi} \ket{\psi(t)} 
   = \sum_j \sum_{\nu,\nu'} c_{\nu,j} (t) c_{\nu',j}^\ast (t) \langle \nu' | {\hat \chi} | \nu \rangle.
\end{equation}
Inserting ${\hat \chi} = \sqrt{\frac{\hbar}{2 m \omega}} ({\hat a} + {\hat a}^\dagger)$, where $m$ is the reduced mass and $\omega = (E_1^{vib}-E_0^{vib})/\hbar$ the normal mode frequency, the expectation value can be written as
\begin{equation}
\langle {\hat \chi} \rangle (t)  = \sqrt{\frac{\hbar}{2 m \omega}} \sum_j \mathfrak{Re} \left \{ \sum_\nu \sqrt{\nu+1} c_{\nu,j}(t) c_{\nu+1,j}^\ast (t)\right \},
   \label{eq:rot_average}
\end{equation}
where we assume that the relevant vibrational states can be approximated by a harmonic oscillator wavefunction. The expectation value is non-zero only if the ro-vibrational states 
with $\nu$ and $\nu+1$ belong to the same rotational state $j$. 
For an ensemble of molecules, a non-zero expectation value $\langle {\hat \chi} \rangle (t) \neq 0$ means that all molecules vibrate in phase despite their random orientation in space, forming a coherent vibrational wavepacket.
The coupling to the  rotational motion thus imposes additional conditions for the excitation of coherent vibrational motion if the molecules are randomly oriented. 

\begin{figure}[t]
	\centering
	\includegraphics[width=8cm]{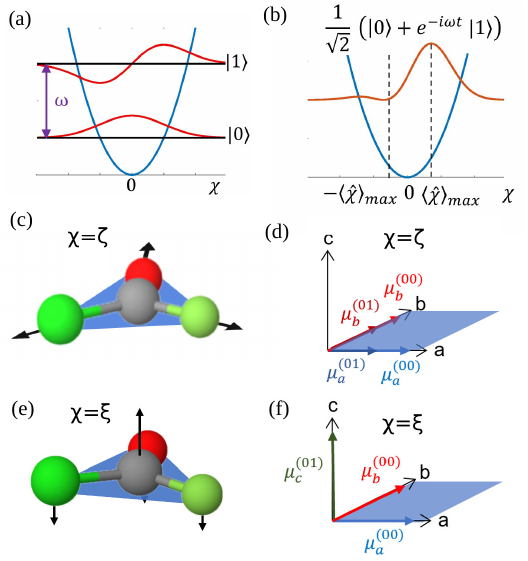}
	\caption{Creation of a vibrational wavepacket in a planar molecule for the example of COFCl. (a) Vibrational eigenstates $\ket{0}$ and $\ket{1}$ of a normal mode $\chi$. (b) Vibrational wavepacket as superposition of eigenstates $\ket{0}$ and $\ket{1}$. (c) Sketch of an in-plane normal mode $\chi = \zeta$ of COFCl (d) Permanent dipole moment of COFCl with components $\mu_a^{(00)}$ and $\mu_b^{(00)}$ lying in the molecular plane. For the in-plane vibration, the transition dipole moment with components $\mu_a^{(01)}$ and $\mu_b^{(01)}$ lies also in the molecular plane. (e) Sketch of the out-of plane normal mode $\chi= \xi$ of COFCl. (f)  Permanent dipole moment with components $\mu_a^{(00)}$ and $\mu_b^{(00)}$ lying in the molecular plane. The transition dipole moment for the out-of plane vibration with component $\mu_c^{(01)}$ perpendicular to the molecular plane.}
	\label{fig:chiral_wavepacket}
\end{figure}

For simplicity, we consider in the following a vibrational wavepacket that consists of the ground and first excited vibrational states only, as shown in Fig.~\ref{fig:chiral_wavepacket}(a). {\color{black} This implies a slight anharmonicity in the vibrational potential so that the energy gaps between subsequent vibrational states are not the same and it is thus possible to address the transition between $\ket{0}$ and $\ket{1}$ without driving the transition between $\ket{1}$ and $\ket{2}$ and so on.}
 In this case, the expectation value for the elongation becomes
 \begin{equation}
 \langle {\hat \chi} \rangle (t) = \sqrt{\frac{\hbar}{2 m \omega}} \sum_j \mathfrak{Re} \left [ b_{0,j}(t) b_{1,j}^\ast (t) \exp (i \omega t) \right ]
 \label{eq:rot_average_01}
 \end{equation}
 where $b_{\nu,j}(t) = c_{\nu,j}(t) \exp [i (E_\nu^{vib} + E_j^{rot})t/\hbar]$. The elongation along the normal mode becomes maximal with $\langle {\hat \chi} \rangle_{max}=\frac{1}{2} \sqrt{\frac{\hbar}{2m \omega}}$ if the population is equally distributed between the ground and excited vibrational state for each rotational state $j$.

In this paper, we discuss the excitation of vibrational wavepackets in a planar molecule, for example COFCl, see Fig.~\ref{fig:chiral_wavepacket}, where the molecular plane is the only symmetry element. 
The normal mode can either describe an in-plane vibration, which we denote by $\chi=\zeta$, see Fig.~\ref{fig:chiral_wavepacket} (c) or an out-of-plane vibration $\chi=\xi$, as shown in Fig.~\ref{fig:chiral_wavepacket} (e). 
Excitation of an in-plane vibrational wavepacket leaves a planar molecule achiral.
Vibrations along the out-of-plane normal mode $\xi$ break the planar symmetry  \cite{Tikhonov22}.
 If a wavepacket with $\langle {\hat \xi} \rangle (t)\neq 0$ is excited, the molecule becomes temporarily chiral and oscillates between its two enantiomeric structures with the frequency of the out-of-plane vibration. 

 In the following, we discuss different interaction schemes for the excitation of coherent vibrational wavepackets in randomly oriented molecules. 
 In Section~\ref{subsec:symmetry}, we make use of symmetry properties of asymmetric top rotors to derive conditions for the external fields ${\bf E}_i(t)$ so that after excitation $\langle {\hat \chi} \rangle (t) \neq 0$, i.e. the average elongation measured for an ensemble of molecules does not vanish. This rather technical part is followed by Section~\ref{subsec:examples}, where the results of Section~\ref{subsec:symmetry} are illustrated with examples for the creation of achiral and chiral wavepackets in planar molecules. In Section~\ref{subsec:graph-method} we address the question of full controllability of the Schr\"odinger equation (\ref{eq:TDSE}) as means of defining conditions for external fields that can induce maximal molecular response $\langle {\hat \chi} \rangle = \langle {\hat \chi} \rangle_{max}$ for any given initial condition. In Sections~\ref{sec:example1} and \ref{sec:example2} we then concentrate on inducing chirality in an achiral molecule and present two examples for the excitation of chiral vibrational wavepackets. 

\section{Conditions for creating a coherent vibrational wavepacket}\label{sec:condition}

\subsection{Symmetry of the ro-vibrational wavefunctions} \label{subsec:symmetry}
According to Eq.~(\ref{eq:rot_average_01}), for the simplest case of two vibrational states forming a coherent superposition, a net chiral signal after averaging over rotations is obtained if the coefficients $b_{0,j}(t)$ and $b_{1,j}(t)$ are non-zero. 
Vanishing of these coefficients due to symmetry arguments can be determined by using perturbation theory to solve the time-dependent Schr\"odinger equation (\ref{eq:TDSE}). Starting from the initial state $\ket{\psi (0)}=\ket{\nu_0} \ket{\phi_{j_0}}$, the coefficients $b_{\nu,j}(t)$ are not identically zero only if 
\begin{widetext}
 \begin{eqnarray}
\bra{\phi_j}\bra{\nu} H_{int}^l \ket{\nu_0} \ket{\phi_{j_0}} =  
\bra{\phi_j}
\bra{\nu} H_{int} \ket{\nu^{(l-1)}} \ket{\phi_{j^{(l-1)}}} \dots \bra{\phi_{j''}} \bra{\nu''} H_{int} \ket{\nu'} \ket{\phi_{j'}} \bra{\phi_{j'}} \bra{\nu'} H_{int} \ket{\nu_0} \ket{\phi_{j_0}} \neq 0 
 \label{eq:trans_matrix_element_exp}
 \end{eqnarray}
 \end{widetext}
 for at least one $l=0,1,2,...$, where $l$ is the order of perturbation and $\ket{\phi_{j'}} \ket{\nu'}$, $\ket{\phi_{j''}} \ket{\nu''}$, and so on, are arbitrary intermediate ro-vibrational states. 
Since a non-zero expectation value $\langle {\hat \chi} \rangle (t)$ in Eq.~(\ref{eq:rot_average_01}) requires both $b_{0,j}(t)$ and $b_{1,j}(t)$ to be non-zero, the following condition has to be fulfilled: For a given initial rotational state $\ket{\phi_{j_0}}$ and $\ket{\nu_0} = \ket{0}$, at least one $j$ must exist for which both transition matrix elements
\begin{subequations}\label{eq:cond_wp}
     \begin{align}
     \bra{\phi_j} \bra{0} H_{int}^l \ket{0} \ket{\phi_{j_0}} &\neq 0\,,&
\label{eq:cond_wp_1} \\
\bra{\phi_j} \bra{1} H_{int}^{l'} \ket{0} \ket{\phi_{j_0}} &\neq 0&
\label{eq:cond_wp_2}
     \end{align}
\end{subequations}
are non-zero for at least one $l$ and $l'$. 

Next, we analyze what Eq.~\eqref{eq:cond_wp} implies for vibrating
asymmetric top rotors. We represent the rotational part of the ro-vibrational wavefunction in the basis of asymmetric top eigenfunctions $\ket{\phi_j}=\ket{J_{K_a,K_c},M}$ with  $J=0,1,2,\dots$ the rotational quantum number,  $M=-J,-J+1,\dots,J$ the projection quantum number for rotation around the space-fixed axis, and $K_a=0,1,\dots,J$ ($K_c=0,1,\dots,J$) the projection quantum number for rotation around the molecular axis of a prolate (oblate) symmetric top~\footnote{Each asymmetric top eigenfunction is uniquely described by $J$, $M$ and the two corresponding symmetric top quantum numbers $K_a$ and $K_c$. Since the rotational energy eigenvalues $E_j^{rot} =E_{J_{K_a,K_c}}^{rot}$ do not depend on $M$, we denote the rotational energy levels by ${J_{K_a,K_c}}$.}.
The symmetry of the ro-vibrational wavefunctions with respect to the space-fixed and molecule-fixed coordinate systems determine whether the transitions matrix elements in Eq.~(\ref{eq:cond_wp}) are non-zero. Symmetry with respect to the space-fixed frame results in the (usual) $M$-selection rules for electric dipole interaction: $\Delta M = 0$  for transitions induced by $z$-polarized fields, and 
$\Delta M = \pm 1$ for transitions induced by $x$- or $y$-polarized fields. Since both conditions in Eq.~(\ref{eq:cond_wp}) need to be fulfilled, the polarizations ${\bf e}_i$ of the external fields ${\bf E}_i(t)$ have to be chosen such that for the $l$th order process in Eq.~(\ref{eq:cond_wp_1}) the final quantum number $M$ is the same as for the $l'$th order process in Eq.~(\ref{eq:cond_wp_2}). We will discuss this condition for several examples in Subsection \ref{subsec:examples}.

Equation~\eqref{eq:cond_wp} also involves the symmetry of the ro-vibrational wavefunctions with respect to the molecule fixed frame. The corresponding condition can be obtained from the properties of the symmetry group $D_2$ of an asymmetric top, recalled in Table~\ref{D2} \cite{Bunker}:
 \begin{table}[t]
 	\vspace{5mm}
 	\begin{tabular}{c|cccc|c}
 		$\,D_2\,$   & $\,E\,$ & $\,R_a^\pi\,$ & $\,R_b^\pi\,$ & $\,R_c^\pi\,$ & $\,K_a K_c\,$ \\
 		\hline
 		$A$  &   1  &   1    &   1 & 1 & ee    \\
 		$B_a$   & 1  &   1    &   -1 & -1 & eo   \\
 		$B_b$   & 1  &   -1    &   1 & -1 & oo   \\
 		$B_c$   & 1  &   -1    &   -1 & 1 & oe   \\
 	\end{tabular}
 	\caption{Character table of $D_2$, the molecular rotation group for asymmetric top molecules, and transformation properties of the asymmetric top eigenfunctions~\cite{Bunker}. 
 		The transformation properties of the 
 		rotational states depend on whether the quantum numbers $K_a$ and $K_c$ are even~(e) or odd~(o).} 
 	\label{D2}
 \end{table}	
 The transition matrix elements, Eq.~(\ref{eq:cond_wp}), can be non-zero only if they transform according to the totally symmetric irreducible representation of $D_2$, i.e,
 \begin{widetext}
   \begin{subequations} \label{eq:sym_cond}
      \begin{align} 
       \Gamma (\ket{\phi_j}) \times \Gamma(\bra{0} H_{int} \ket{\nu^{(l-1)}} ) \times 
       ... \times
       \Gamma(\bra{\nu''} H_{int} \ket{\nu'}) \times 
       \Gamma(\bra{\nu'} H_{int} \ket{0}) \times \Gamma(\ket{\phi_{j_0}}) = A 
       \label{eq:sym_cond_1} \\
       \Gamma (\ket{\phi_j}) \times \Gamma(\bra{1} H_{int} \ket{\lambda^{(l'-1)}} ) \times 
       ... \times
       \Gamma(\bra{\lambda''} H_{int} \ket{\lambda'}) \times 
       \Gamma(\bra{\lambda'} H_{int} \ket{0}) \times \Gamma(\ket{\phi_{j_0}}) = A
       \label{eq:sym_cond_2}
       \end{align}
   \end{subequations}  
 \end{widetext}
 for conditions (\ref{eq:cond_wp_1}) and (\ref{eq:cond_wp_2}), respectively, 
 where $\nu',...,\nu^{(l-1)}, \lambda',...,\lambda^{(l'-1)} \in\{0,1\}$.
 According to Table~\ref{D2}, the irreducible representations of the rotational wavefuctions are
 $\Gamma(\ket{\phi_j}) = \Gamma(\ket{J_{K_a,K_c},M}) = A$ or $B_\alpha$, with $\alpha=a,b,c$, depending only on
 the values of $K_a$ and $K_c$. In Eq.~(\ref{eq:sym_cond}), we have further utilized that $B_\alpha \times B_\alpha = A$. 

 Finally, Eq.~(\ref{eq:cond_wp}) needs to be evaluated for the vibrational part. 
 In order to determine the irreducible representations of the vibrational transition matrix elements in Eq.~(\ref{eq:cond_wp}), we decompose the interaction Hamiltonian into its irreducible components.
 Therefore, we consider an electric field ${\bf E}= {\bf e}_p {\cal E} u(t)$ with polarization ${\bf e}_p$, amplitude ${\cal E}$ and time-dependence $u(t)$. The vibrational transition matrix elements can be written as
 \begin{widetext}
 \begin{equation}
  \bra{\nu'} H_{int}\ket{\nu} = \sum_\alpha \bra{\nu'} H_{int,\alpha} \ket{\nu} 
  = - {\cal E} u(t) \sum_\alpha \mu_\alpha^{(\nu'\nu)} R_{\alpha,p}(\gamma_R)
  \label{eq:Hint_alpha}
 \end{equation}
 \end{widetext}
 with $\alpha=a,b,c$ and $\mu_\alpha^{(\nu \nu')} = \bra{\nu'} \mu_\alpha \ket{\nu}$. Here, $\mu_\alpha^{(01)}=\mu_\alpha^{(10)}$ are transition dipole moments. For the permanent dipole moments we assume for simplicity that $\mu_\alpha^{(11)}=\mu_\alpha^{(00)}$. The matrix elements of the rotational matrix ${\bm R}(\gamma_R)$ are denoted by $R_{\alpha,p} (\gamma_R)$.
The irreducible components of the interaction Hamiltonian transform according to the irreducible representations of $D_2$, Table~\ref{D2}, namely
 \begin{subequations}\label{eq:sym_Hint}
     \begin{eqnarray}
 \bra{\nu'} H_{int,a} \ket{\nu} &\sim& B_a \,, \\
 \bra{\nu'} H_{int,b} \ket{\nu} &\sim& B_b \,, \\
 \bra{\nu'} H_{int,c} \ket{\nu} &\sim& B_c\,.
 \end{eqnarray}  
 \end{subequations}
In order to obtain non-vanishing coefficients $b_{0,j}(t)$ and $b_{1,j}(t)$ 
the excitation process has to contain an $l$th order process with
\begin{equation}
     \Gamma \left (\bra{0} H_{int} \ket{\nu^{(l-1)}} \right) \times ...
   \times
    \Gamma \left (\bra{\nu'} H_{int} \ket{0} \right) = \Gamma_0,
   \label{eq:cond_dip1} 
\end{equation}
where $\Gamma_0$ is one of the irreducible representations of $D_2$, and an $l'$th order process that fulfills
\begin{equation}
  \Gamma \left (\bra{1} H_{int} \ket{\lambda^{(l'-1)}} \right) \times ... \times 
    \Gamma \left (\bra{\lambda'} H_{int} \ket{0} \right) = \Gamma_0
     \label{eq:cond_dip2} 
\end{equation}
with the same $\Gamma_0$. 
Equations (\ref{eq:sym_cond_1}) and (\ref{eq:sym_cond_2}) are thus both fulfilled if
$\Gamma (\ket{\phi_j}) \times \Gamma_0 \times \Gamma (\ket{\phi_0}) = A$, i.e.,
the final rotational state $\ket{\phi_j}$ transforms according to $\Gamma (\ket{\phi_j}) = \Gamma( \ket{\phi_0}) \times \Gamma_0$.
In combination with Eq.~\eqref{eq:sym_Hint}, the conditions~(\ref{eq:cond_dip1})  and (\ref{eq:cond_dip2}) can be utilized to determine which combination of external fields ${\bf E}_i$ is capable to excite a coherent vibrational wavepacket in initially randomly oriented molecules.

 \subsection{Examples of creating coherent ro-vibrational wavepackets} \label{subsec:examples}
 We now show how to use the conditions just derived to create coherent vibrational wavepackets, both achiral and chiral. To this end, we
 consider a planar molecule, with the molecular plane as the only symmetry element, e.g. COFCl, as shown in Fig.~\ref{fig:chiral_wavepacket}. Due to the planar symmetry, the molecule has a permanent dipole moment in the molecular plane, i.e., $\mu^{(00)}_a \neq 0$ and $\mu^{(00)}_b \neq 0$ while $\mu^{(00)}_c = 0$. 
 For an in-plane vibration $\chi=\zeta$, also the transition dipole moment lies in the molecular plane, i.e $\mu_a^{(01)} \neq 0$, $\mu_b^{(01)} \neq 0$ and $\mu_c^{(01)} = 0$, see Fig.~\ref{fig:chiral_wavepacket}(d). For an out-of-plane vibration $\chi=\xi$, the transition dipole moment is perpendicular to the molecular plane, i.e $\mu_a^{(01)} = 0$, $\mu_b^{(01)} = 0$ and $\mu_c^{(01)} \neq 0$, as in Fig.~\ref{fig:chiral_wavepacket}(e), see also Ref.~\cite{Tikhonov22}. In the following, we discuss various excitation scenarios, assuming that the molecules are initially in their vibrational and rotational ground state $\ket{0} \ket{\phi_0} = \ket{0} \ket{0_{0,0},0}$. 
 
 (a) Ro-vibrational excitation in first order perturbation theory, i.e., excitation of a vibrational state with one IR-photon is described by the case $l=0$ and $l'=1$. This reduces condition (\ref{eq:cond_wp_1}) to 
 $\langle \phi_j |  \phi_{j_0} \rangle \neq 0$, which requires $j=j_0$. For $l'=1$ and $j=j_0$, condition~\eqref{eq:cond_wp_2} can only be fulfilled if $\Gamma(\bra{1} H_{int} \ket{0})=A$. However, this is a contradiction to Eq.~(\ref{eq:sym_Hint}). It is thus not possible to excite a vibrational wavepacket with $\langle {\hat \chi} \rangle \neq 0$ with a single IR interaction in a sample of randomly oriented molecules. The same result is obtained when the molecular rotation is treated classically \cite{Tikhonov22}. In both cases, it corresponds at any instant of time to as many molecules (each with a given orientation) with $+\langle\hat{\chi}\rangle$ as with $-\langle\hat{\chi}\rangle$ such that the overall normal mode elongation vanishes when averaged over random orientations.

 (b) Next, we consider the case $l=1$ and $l'=1$. 
 Conditions (\ref{eq:cond_dip1})  and (\ref{eq:cond_dip2}) are then fulfilled if $\mu_\alpha^{(01)} \neq 0$ and $\mu_\alpha^{(00)} \neq 0$ for the same $\alpha$, i.e., if both the transition dipole moment and the permanent dipole moment have a non-vanishing component along the same molecular axis. This is the case if the normal mode is an in-plane vibration $\chi=\zeta$ with $\mu_a^{(00)}$ and $\mu_a^{(01)}$ or with $\mu_b^{(00)}$ and $\mu_b^{(01)}$, see Fig.~\ref{fig:chiral_wavepacket}(d). A purely rotational transition within the vibrational ground state with 
 $
 \bra{0} H_{int} \ket{0} 
 \neq 0$ can be realized with a microwave field resonant to an allowed rotational transition. The vibrational transition with $
 \bra{1} H_{int} \ket{0}
 \neq 0$ can be driven by an IR-pulse resonant to the frequency of the normal mode.
In addition, the overall excitation must fulfill the $M$-selection rules. Starting from the rotational state $\ket{\phi_{0}}=\ket{0_{0,0},0}$, the transitions described by Eq.~(\ref{eq:cond_wp}), namely $\bra{J_{K_a,K_c},M} \bra{0} H_{int} \ket{0} \ket{0_{0,0},0}$ and $\bra{J_{K_a,K_c},M} \bra{1} H_{int} \ket{0} \ket{0_{0,0},0}$, must both end in the same rotational state $M$. Since one can always chose the quantization axis, we can assume, without loss of generality, the first interaction to be with a $z$-polarized field, inducing rotational transitions with $\Delta M=0$. Then $M=0$, and the second field also has to be $z$-polarized such that condition (\ref{eq:cond_wp_2}) is fulfilled with $M=0$.
Thus, an achiral vibrational wavepacket with $\langle {\hat \zeta} \rangle \neq 0$ can be excited with a combination of a $z$-polarized microwave pulse and an IR-pulse with the same polarization. However, excitation of a chiral vibrational wavepacket is not possible.

 (c) In order to induce a chiral vibrational wavepacket, it is necessary to excite the out-of plane vibrational mode $\chi=\xi$. In this case, the transition dipole moment is perpendicular to the permanent dipole moment, see Fig.~\ref{fig:chiral_wavepacket}(f). The lowest order for which   conditions~\eqref{eq:cond_dip1}  and (\ref{eq:cond_dip2}) can be fulfilled in this case is for $l=2$ and $l'=1$ (or vice versa) with 
 \begin{equation}
   \Gamma (\bra{0} H_{int,a} \ket{0} ) \times
   \Gamma (\bra{0} H_{int,b} \ket{0} ) = B_a \times B_b = B_c
   \label{eq:sym_cond_MW}
 \end{equation}
and
\begin{equation}
   \Gamma (\bra{1} H_{int,c} \ket{0} ) = B_c\,.
    \label{eq:sym_cond_IR}
 \end{equation}
In other words, three interactions are required, transforming according to the irreducible representations $B_a$, $B_b$ and $B_c$, respectively. Equation~\eqref{eq:sym_cond_MW} describes purely rotational transitions within the vibrational ground state which can be realized with two microwave pulses, while the vibrational transition Eq.~(\ref{eq:sym_cond_IR}) can be driven by an IR-pulse. We discuss such an excitation scheme in Section~\ref{sec:example1}.
Exchanging the roles of $l$ and $l'$, condition~(\ref{eq:cond_dip1}) can also be fulfilled by
  \begin{equation}
   \Gamma (\bra{0} H_{int,a} \ket{1} ) \times
   \Gamma (\bra{1} H_{int,b} \ket{0} ) = B_a \times B_b = B_c.
   \label{eq:sym_cond_stat}
  \end{equation}
  Equation~(\ref{eq:sym_cond_stat}) together with Eq.~(\ref{eq:sym_cond_IR}) describes the conditions for exciting a chiral wavepacket with three ro-vibrational transitions, which can be driven by three IR-pulses.
  For a planar molecule in free space, $\bra{0} H_{int,\alpha} \ket{1} = 0$ for $\alpha=a,b$. In Section~\ref{sec:example2} we show that 
  such transition matrix elements do occur in the presence of an external static electric field.

 Moreover, the above conditions imply that the three electric fields that are (at least) necessary to excite a chiral vibrational wavepacket, have to be polarized orthogonal to each other. Starting with $M_0=0$ and assuming without loss of generality that the first order transition  $\bra{\phi_j}\bra{1} H_{int,c} \ket{0}\ket{\phi_0}$ is driven by $z$-polarized field, the final rotational state has the quantum numbers $J=J_0+1=1$ and $M=M_0=0$. This rotational state can be addressed by a second order process only with a combination of $x$- and $y$-polarized fields \cite{Leibscher19}.
 Note that this corresponds exactly to the general condition for distinguishing enantiomers when exciting molecules that are chiral in the first place \cite{OrdonezPRA18}. It has also been determined as condition for creating a chiral wavepacket when considering a classical angular distribution of the rotors \cite{Tikhonov22}. 

\subsection{Controllability of the ro-vibrational 
Schr\"odinger equation}\label{subsec:graph-method}
The conditions discussed in Sec.~\ref{subsec:symmetry} and \ref{subsec:examples} ensure that a coherent vibrational wavepacket can be created starting from the ground ro-vibrational state. In order to ensure how a coherent vibrational wavepacket with \textit{maximal} elongation $\langle {\hat \chi} \rangle_{max}$ can be excited from an \textit{arbitrary} initial state, we pursue a 
different approach, namely we analyze the controllability \cite{Alessandro2008}. 
{\color{black}
The Schr\"odinger equation is 
said to be 
controllable 
if the system can be steered from any given initial condition to any given target state, with a suitable choice of control fields ${\bf E}_i(t)$, $i=1,\dots,f$.
}
This implies that one can create a target state corresponding to maximal elongation $\langle \hat{\chi} \rangle$. 

In order to analyze the controllability of the ro-vibrational Schr\"odinger equation (\ref{eq:TDSE}) we make use of graph theory based methods \cite{BCCS,BCS,Gago_2023}. We consider a graph $\mathcal{G}$ consisting of the eigenstates $\ket{\nu} \ket{\phi_j}$ of $H_0$ as nodes and the non-zero transition matrix elements 
$\bra{\phi_j} \bra{\nu} H_{int} \ket{\nu'} \ket{\phi_{j'}}$ as edges. A quantum system has been shown to be 
controllable  if the associated graph $\mathcal{G}$ has a connected sub-graph that contains all nodes of $\mathcal{G}$ and only decoupled transitions \cite{CMSB}, see also \footnote{{\color{black} Actually, this condition implies a stronger notion of controllability, namely, controllability at the level of the propagators, which implies, in particular, controllability of the density matrices.}}.  
{\color{black} Here, two transitions are called coupled by a control Hamiltonian if they are non-vanishing and have equal energy gaps.
A transition is uncoupled, if for any other transition, there exists at least one control Hamiltonian that does not couple them. Finally, a transition is decoupled, if for any other transition, there exists at least one nested commutator (of arbitrary length) between control and drift Hamiltonians
that does not couple them \cite{Gago_2023}. 
For a quantum rotor, most transitions are coupled due to the degeneracy of the rotational states.} 
Recently, a graph-theoretical method to decouple the resonant transitions between asymmetric top states has been developed \cite{Pozzoli21}, and the maximal number of external fields required to control finite subsystems of an asymmetric top have been identified  \cite{Leibscher22}. Here, we generalize these methods to analyze the controllability of the \textit{ro-vibrational} Schr\"odinger equation (\ref{eq:TDSE}).
However, controllability analysis requires an {\it a priori} selection of controls, i.e., external fields that interact with the molecule, in contrast to symmetry analysis. We carry out controllability analysis for two practical examples in Sec.~\ref{sec:example1} and \ref{sec:example2},
focusing on the excitation of a chiral vibrational wavepacket under the conditions derived in \ref{subsec:examples}(c). 
In particular, in Section~\ref{sec:example1}, we study the interaction of a planar molecule with a combination of microwave pulses and an IR pulse, whereas in Section~\ref{sec:example2}, we consider the interaction with a sequence of IR-pulses in the presence of a static electric field. We  exemplify in both cases that the symmetry conditions derived in \ref{subsec:examples} (c) are fulfilled and then prove  controllability to ascertain that the excitation process can result in a \textit{maximal} chiral response, irrespective of the initial state.

\section{Excitation with a combination of microwave and IR pulses}\label{sec:example1}
In the first example, microwave pulses are employed to induce transitions between rotational states of an asymmetric top molecule and IR radiation is applied to produce an excited vibrational state of the out-of-plane normal mode. 

The Hamiltonian describing the interaction with microwave and IR-pulses in electric dipole approximation reads
\begin{equation}
H_{int}(t) = H_{int}^{MW}(t) + H_{int}^{IR}(t).
\label{eq:Hamiltonian_MW_IR}
\end{equation}
The interaction of the microwave pulses with the electric fields ${\bf E}_i(t)$ as defined in Eq.~(\ref{eq:electric_fields})
is then given by 
\begin{equation}
H_{int}^{MW}(t) =  \sum_i u_i(t) H_i^{MW}
\end{equation}
with
\begin{widetext}
\begin{equation}
  H_i^{MW} = -{\cal E}_i   \sum_{j,j'} \sum_{\nu=0}^{1} \ket{\nu} \ket{\phi_{j'}} \bra{\phi_{j'}} \left( {\bm \mu}^{(00)} \cdot {\cal R}(\gamma_R) \cdot {\bm e}_i \right) \ket{\phi_{j}} \bra{\phi_{j}} \bra{\nu} + c.c.
  \label{eq:Hint_MW}
\end{equation}
 The transformation between the space fixed and molecule fixed coordinate system can be expressed in terms of
 the Wigner D-matrix elements  $D_{MK}^J(\gamma_R)$ ~\cite{Zare88} as
	\begin{eqnarray}
	{\bm \mu}^{(00)} \cdot {\cal R}(\gamma_R) \cdot {\bm e}_{x}  &=& \frac{\mu_a^{(00)} }{\sqrt{2}}  \left( D_{-10}^1  - D_{10}^1 \right)  + \frac{\mu_b^{(00)} }{ 2} \left ( D_{11}^1 - D_{1-1}^1  - D_{-11}^1 + D_{-1-1}^1\right) \nonumber \\
	{\bm \mu}^{(00)} \cdot {\cal R}(\gamma_R) \cdot {\bm e}_{y}  &=& -\mathrm{i} \frac{\mu_a^{(00)} }{\sqrt{2}}  \left( D_{-10}^1  + D_{10}^1 \right)  + \mathrm{i} \frac{\mu_b^{(00)} }{ 2} \left ( D_{11}^1 - D_{1-1}^1  + D_{-11}^1 - D_{-1-1}^1\right), \nonumber \\
	{\bm \mu}^{(00)} \cdot {\cal R}(\gamma_R) \cdot {\bm e}_{z}  &=&  \mu_a^{(00)}  D_{00}^1 - \frac{\mu_b^{(00)} }{\sqrt{2} } \left ( D_{01}^1 - D_{0-1}^1 \right).
	\label{mu_projection_00}
	\end{eqnarray}  
\end{widetext}
 The evaluation of the matrix elements $\bra{\phi_{j'}} {\bm \mu}^{(00)} {\cal R}(\gamma_R) {\bm e}_{p} \ket{\phi_{j}}$ between the asymmetric top eigenstates $\ket{\phi_{j}} = \ket{J_{K_a,K_c},M}$ is described e.g. in \cite{Leibscher19,Leibscher22}. 
 The time-dependence of the microwave pulses can be written as $u_i(t) = s_i(t) \cos (\omega_i t + \phi_i)$, where $s_i(t)$ is the dimensionless envelope and $\omega_i$ and $\phi_i$ are frequency and phase of the field.
The frequencies $\omega_i$ are chosen to be resonant to one of the rotational transitions. The field intensity can then be tuned such that only those transitions resonant to the frequency of the field are excited \cite{Leibscher22}, and $\bra{\phi_{j'}}  {\bm \mu}^{(00)} \cdot {\cal R}(\gamma_R) \cdot {\bm e}_i \ket{\phi_{j}} = 0$ if $\omega_i \neq |E_j^{rot} - E_{j'}^{rot}|/\hbar$.

The frequency of the IR pulse is chosen such that it is (approximately) resonant to the transition between the vibrational states $\ket{0}$ and $\ket{1}$. That is, only transitions between the vibrational states can occur, but no rotational transitions within one of the vibrational states. The corresponding interaction Hamiltonian can thus be expressed as
\begin{equation}
H_{int}^{IR}(t) =  u(t) \hat{H}^{IR}, 
\end{equation}
with
\begin{widetext}
\begin{equation}
\hat{H}^{IR} = - {\cal E}_{IR} \sum_{j,j'} \ket{0} \ket{\phi_{j'}} \bra{\phi_{j'}}  \left ( {\bm \mu}^{(01)} \cdot {\cal R}(\gamma_R) \cdot {\bm e}_{IR} \right) \ket{\phi_{j}} \bra{\phi_{j}} \bra{1} + c.c.,
\end{equation}
\end{widetext}
where ${\bm e}_{IR}$ denotes the polarization of the IR pulse. Recall that the transition dipole moment has only one component perpendicular to the molecular plane i.e. only $\mu_c^{(01)} \neq 0$. The transformation from the space fixed to the molecule fixed frame is thus given by \cite{Zare88}
\begin{eqnarray}
{\bm \mu}^{(01)} {\cal R}(\gamma_R) {\bm e}_{x}  &=& -  \mathrm{i} \frac{\mu_c^{(01)}}{2} \left ( D_{11}^1 + D_{1-1}^1  - D_{-11}^1 - D_{-1-1}^1\right), \nonumber \\
{\bm \mu}^{(01)} {\cal R}(\gamma_R) {\bm e}_{y}  &=&  \frac{\mu_c^{(01)}}{2} \left ( D_{11}^1 + D_{1-1}^1  + D_{-11}^1 + D_{-1-1}^1\right), \nonumber \\
{\bm \mu}^{(01)} {\cal R}(\gamma_R) {\bm e}_{z}   &=&  \mathrm{i} \frac{\mu_c^{(01)}}{\sqrt{2}} \left ( D_{01}^1 + D_{0-1}^1 \right).
\label{mu_projection_01}
\end{eqnarray}  
The time-dependence of the IR-pulse can be written as $u(t) = s(t) \cos (\omega_{IR} t + \phi)$. If the duration of the IR-pulse is longer that the rotational period of the molecule, its bandwidth is small enough to selectively excite individual ro-vibrational states, i.e.
only transitions with $\omega_{IR}=|E_{j'}^{rot}+E_0^{vib} - E_{j}^{rot}-E_1^{vib}|/\hbar$. However, the bandwidth of a short IR pulse can be large enough to excite several rotational states simultaneously. In this case the interaction Hamiltonian contains all transition matrix elements which are allowed according to the dipole selection rules. 

In the following, we show that the combination of microwave and IR-pulses fullfills the requirements derived in \ref{subsec:examples} (c) for the creation of a chiral wavepacket. We analyze the controllability of the ro-vibrational  Schr\"odinger equation (\ref{eq:TDSE}) with an interaction Hamiltonian of the form Eq.~(\ref{eq:Hamiltonian_MW_IR}) and discuss the results of numerical simulations demonstrating the excitation of a chiral wavepacket with a combination of microwave and IR-pulses.

\subsection{Conditions for exciting a chiral wavepacket}
\label{subsec:cond_MR_IR}
The conditions for creating a chiral vibrational wavepacket with a combination of microwave and IR pulses can be deduced directly from \ref{subsec:examples} (c). With the two components of the permanent dipole moment, $\mu_a^{(00)}$ and $\mu_b^{(00)}$, condition Eq.(\ref{eq:sym_cond_MW}) is fulfilled with two microwave pulses, one resonant to a b-type transition, e.g. from level $J_{K_a,K_b}=0_{0,0}$ to  $1_{1,1}$ and the second one resonant to an a-type transition e.g. $1_{1,1} \rightarrow 1_{1,0}$ within the vibrational ground state $\ket{0}$. Condition Eq.(\ref{eq:sym_cond_IR}) is fulfilled since the transition dipole moment has the non-vanishing component $\mu_c^{(01)}$ which drives the transition from $\ket{0} \rightarrow \ket{1}$ and  $0_{0,0} \rightarrow 1_{1,0}$. Moreover, as shown in \ref{subsec:examples} (c), three orthogonal polarization directions are required, e.g. a $z$-polarized IR-pulse and $x$- and $y$-polarized microwave pulses.
 
\subsection{Controllability analysis}\label{subsec:control_MR_IR}
\begin{center}
   \begin{figure}[t]
            \centering
            \includegraphics[width=16cm]{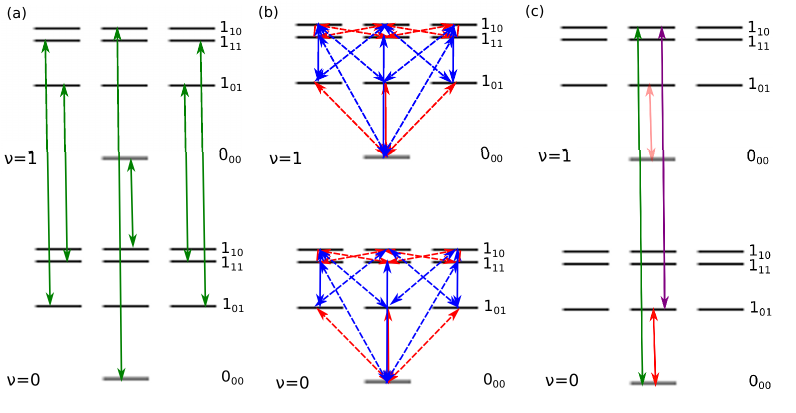}
            \caption{Graph with asymmetric top eigenstates as nodes (horizontal lines) and transition matrix elements as edges (green, red and blue arrows). (a) Interaction with a $z$-polarized IR-pulse (green arrows). (b) Interaction with $x$-, $y$-, and $z$-polarized microwave pulses. Red arrows indicate transition matrix elements $\bra{\phi_j'} \mu_a^{(00)} \ket{\phi_j} $ and blue arrows correspond to
    transition matrix elements  $\bra{\phi_j'} \mu_b^{(00)} \ket{\phi_j}$. (c) Graphical commutator between the transitions indicated by the green and red arrows, allowing the two coupled transitions indicated by the red arrows to decouple.}
	\label{fig:controllability_MW_IR}
        \end{figure}    
    \end{center}

We now demonstrate that the Schr\"odinger equation (\ref{eq:TDSE}) with an interaction Hamiltonian of the form Eq.~(\ref{eq:Hamiltonian_MW_IR}) is  
controllable, implying that maximal elongation can be obtained for an arbitrary initial state. 
 We first consider a subspace of the asymmetric top eigenstates consisting of all rotational states with quantum number  $J=0$ and $J=1$. The corresponding graph is shown in Fig.~\ref{fig:controllability_MW_IR}, where the nodes of the graph, i.e., the ro-vibrational eigenstates $\ket{\nu} \ket{\phi_j}$ are indicated by  horizontal black lines. The green lines in panel (a) show the transition matrix elements (edges) induced by the control Hamiltonian ${\hat H}^{IR}$. It can be seen that the graph is not connected, i.e. it is not possible to reach, from an arbitrary initial state, all other states of the system by following the edges. Thus, the system is not controllable with a single IR-pulse.
The edges corresponding to the microwave pulses Eq.~(\ref{eq:Hint_MW}) are shown in panel (b) where all three polarization directions $x$, $y$ and $z$ are considered. The red and blue lines correspond to transitions driven by $\mu_a^{(00)}$ and  $\mu_b^{(00)}$, respectively. A finite-dimensional subsystem {\color{black} including the eigenstates of at least two consecutive values of $J$ of a single quantum asymmetric top} is controllable with $x$, $y$ and $z$ polarized fields via electric dipole interaction if the rotor exhibits two orthogonal components of the permanent dipole moment \cite{Pozzoli22}. Since COFCl has two orthogonal dipole moment components in the molecular plane, the rotational part of the Hamiltonian (i.e. the lower graph in panel (b)) is controllable. This means that each of the edges (red and blue lines) corresponds to a decoupled transition. The graph of the ro-vibrational system consists of two identical rotational subsystems, shifted by the vibrational energy. By interaction with microwave fields alone, this system is not controllable since the graph is not connected (see panel (b)). Moreover, each transition in the lower rotational subsystem is coupled to the corresponding transition in the upper rotational subsystem. 

In order to show that the ro-vibrational system is controllable with a combination of a ($z$-polarized) IR pulse and $x$-, $y$- and $z$-polarized microwave pulses, we have to show that the combined graph containing all transitions shown in panels (a) and (b) is connected and contains only {\color{black} uncoupled} or decoupled transitions. It is easy to see that the graph is connected if one of the edges in panel (a) is added to the graph shown in panel (b), e.g. the transition between the states $\ket{0} \ket{0_{00},0}$ and $\ket{1} \ket{1_{10},0}$ (green line in panel (c)). The corresponding transition is {\color{black} uncoupled} because there exists no other transition with the same energy gap. What remains is to show that the pairs of coupled transitions (identical transitions in the upper and lower part of panel (b)) can be decoupled. This can be proven by using graphical commutators \cite{Pozzoli21,Gago_2023}.
 Note that the commutator between a $N\times N$ matrix with 
\onecolumngrid
\twocolumngrid
\noindent a single pair of non-zero elements $(n,m)$, $(m,n)$ with a $N\times N$ matrix with non-zero elements $(n,k)$, $(k,n)$ with $n\neq m \neq k$ is a matrix with non-zero elements $(m,k)$, $(k,m)$.
 If the non-zero matrix elements present edges of a graph, the edge $(m,k)$ between the nodes $m$ and $k$ is called the graphical commutator of $(n,m)$ and $(n,k)$ \cite{Gago_2023}. Panel (c) shows the graphical commutator between the (decoupled) transition $\ket{0} \ket{0_{00},0} \leftrightarrow  \ket{1} \ket{1_{10},0}$ (green line) and the pair of coupled transitions $\ket{\nu} \ket{0_{00},0} \leftrightarrow  \ket{\nu} \ket{1_{01},0}$, $\nu=0,1$ (red lines). The resulting commutator is indicated by the purple line. The graphical commutator between the green and purple edges then results in the lower red line, corresponding to the transition $\ket{0} \ket{0_{00},0} \leftrightarrow  \ket{0} \ket{1_{1},0}$. By this procedure, we have shown, that the transition $\ket{0} \ket{0_{00},0} \leftrightarrow  \ket{0} \ket{1_{1},0}$ is decoupled from $\ket{1} \ket{0_{00},0} \leftrightarrow  \ket{1} \ket{1_{01},0}$. Having decoupled one pair of coupled transitions, all other pairs can also become decoupled by again taking (graphical) commutators between a decoupled transition and a pair of coupled transitions. This allows us to conclude that the Schr\"odinger equation for the ro-vibrational interacting with a combination of an IR-pulse with three microwave fields is controllable. 
Note that in Fig.~\ref{fig:controllability_MW_IR} we only show the rotational state with $J=0$ and $J=1$. However, since every finite-dimensional rotational subsystem {\color{black}including the eigenstates of at least two consecutive values of $J$} of an asymmetric top is controllable \cite{Pozzoli22}, the result is also valid for larger rotational subsystems. We have thus shown that the time dependent Schr\"odinger equation with interaction Hamiltonian~(\ref{eq:Hamiltonian_MW_IR}) is controllable for the out-of-plane vibration of COFCl with a combination of a $z$-polarized IR-pulse and three orthogonally polarized microwave pulses, implying that maximal elongation $\langle {\hat \xi} \rangle_{max}$ can be obtained with such a set of pulses for an arbitrary initial state. 

According to Sections~\ref{subsec:examples} and ~\ref{subsec:cond_MR_IR}, two microwave pulses, polarized in $x$- and $y$-direction in combination with a $z$-polarized IR-pulse are already sufficient to obtain a chiral signal, although not necessary the maximal possible, if the initial state is the ro-vibrational ground state. In Section~\ref{subsec:dynamics_MW_IR}, we make use of this simpler condition to numerically demonstrate the excitation of a chiral wavepacket with a combination of microwave and IR pulses. 

\subsection{Chiral ro-vibrational dynamics} \label{subsec:dynamics_MW_IR}
To demonstrate the excitation of a chiral wavepacket with a combination of microwave and IR pulses numerically, 
we consider the simplest initial condition describing randomly oriented rotors, i.e.,
we assume that the molecules are initially in their ground vibrational and rotational state $\ket{\psi(0)} = \ket{0} \ket{0_{00},0}$.  As an example, we consider COFCl molecules \cite{Tikhonov22} with rotational constants $A=11781.84$ MHz, $B=5246.37$ MHz, $C=3627.49$ MHz and dipole moments $\mu_a^{(00)}=-1.1$ D, $\mu_b^{(00)}=0.8$ D.

\begin{figure}[t]
	\centering
    \includegraphics[width=8cm]{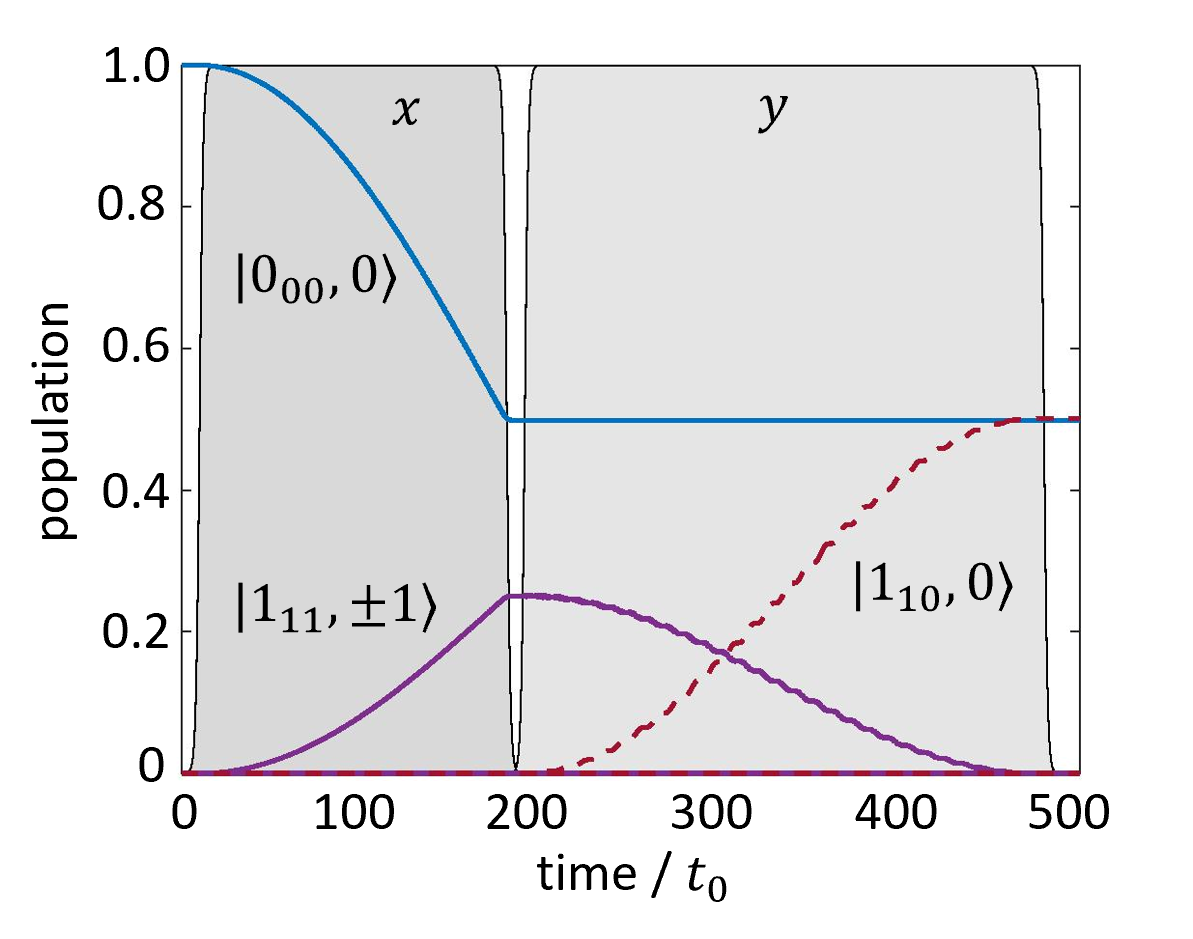}
	\caption{Excitation of a rotational wavepacket in $\nu=0$ with two microwave pulses. Population dynamics for the rotational states $\ket{0_{00},0}$ (solid blue), $\ket{1_{11},\pm 1}$ (magenta) and $\ket{1_{10},0}$ (dashed red).
    The envelope of the two microwave pulses is indicated by the gray shapes, they are $x$- and $y$-polarized as indicated in the figure. Time is given in units of $t_0 = \hbar/B \approx 30$ ps. The amplitude of the microwave fields is ${\cal E}_i = 2 \times 10^4$ V/m for the three fields {i=1,2}.}		
   	\label{fig_population_MW}
\end{figure}

As shown in \ref{subsec:cond_MR_IR}, the excitation of a chiral wavepacket requires two microwave pulses polarized in $x$- and $y$-direction, respectively. The population dynamics in the vibrational ground state during the interaction with microwave pulses, obtained by numerically solving the time-dependent Schr\"odinger equation (\ref{eq:TDSE}) with interaction Hamiltonian (\ref{eq:Hamiltonian_MW_IR}), is shown in Fig.~\ref{fig_population_MW}. 
The duration of the first microwave pulse 
is chosen such that 50 \% of the population remains in the ground state $\ket{0} \ket{0_{00},0}$ while 25 \% is transferred to each of the states $\ket{0} \ket{1_{11},\pm 1}$. The second pulse then completely transfers the population of the states $\ket{0} \ket{1_{11},\pm 1}$ to $\ket{0}\ket{1_{10},0}$, so that the wavefunction after excitation with the microwave-pulses reads
\begin{equation}
\ket{\psi(t)} = \frac{1}{\sqrt{2}} \left ( \ket{0_{00},0} + \exp [i \phi(t)]  \ket{1_{10},0} \right) \ket{0},
\label{eq:psi_MW}
\end{equation}
with the dynamical phase $\phi(t) = - (E_{1_{10}} - E_{0_{00}}) t/\hbar +  \tilde{\phi}$, where $E_{J_{K_a,K_c}}$ are the energies of the asymmetric top eigenstates and $\tilde{\phi}=\phi_2-\phi_1$ is the relative phase between the microwave fields. 
The vibrational state $\ket{1}$ is excited by a $z$-polarized IR-pulse. 
\begin{figure}[t]
	\centering%
	\includegraphics[width=8cm]{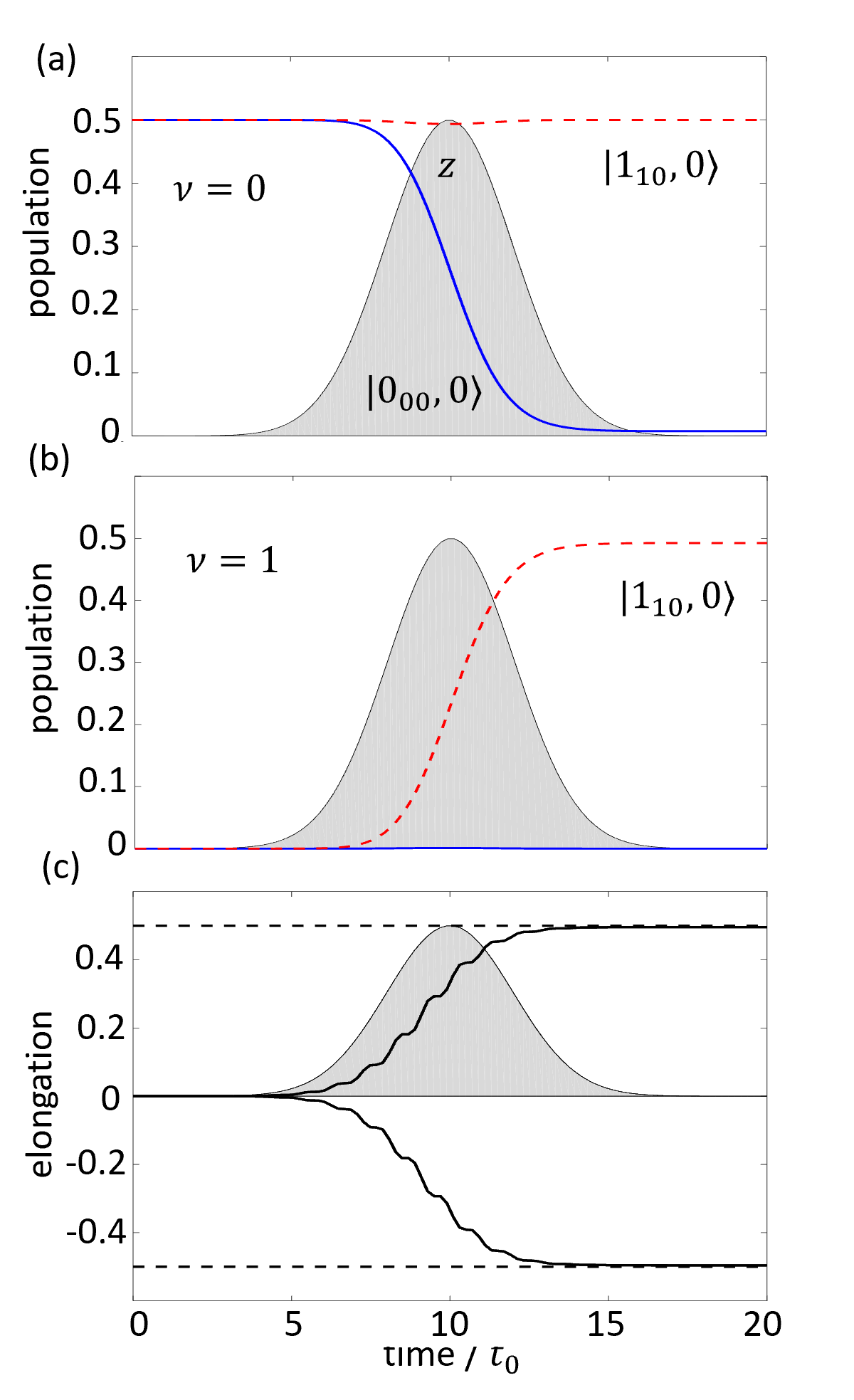}
	\caption{Population dynamics after excitation with a long $z$-polarized IR-pulse. Panels (a) and (b) show the population in the ground and excited vibrational states, respectively. The solid blue and dashed red lines correspond to the rotational states $\ket{0_{00},0}$ and $\ket{1_{10},0}$, respectively. (c) Envelope of elongation $\langle \hat{\xi} \rangle (t) / \xi_0$ during the excitation with the IR-pulse. (The oscillation of $\langle \hat{\xi} \rangle (t)$ with $T=0.01 t_0$ is too fast compared to the timescale of the plot to be resolved.)  The maximal elongation $\pm \langle \hat{\xi}\rangle_{max} / \xi_0$ is indicated by the horizontal dashed lines.  In all panels, the envelope of the IR-pulse is indicated by the gray shapes, the maximal amplitude is ${\cal E}_{IR}=5\times 10^5$ V/m.}
	\label{fig:population_MW_IR_long}
\end{figure}
Figure~\ref{fig:population_MW_IR_long} displays the excitation with a narrowband IR pulse with
\begin{equation}
u(t) = \exp \left(- \frac{(t-t_0)^2}{2 \Delta t^2 }\right) \cos  (\omega_L t + \phi),
\end{equation}
$\omega_L = \omega + E_{1_{10}} / \hbar$ and a bandwidth $\Delta \omega = 1/\Delta t$ small enough to excite only the transition from $\ket{0_{00},0} \ket{0}$ to $\ket{1_{10},0} \ket{1}$. The resulting population dynamics is shown in Fig.~\ref{fig:population_MW_IR_long}(a) and (b). Panel (c) shows the envelope of the elongation $\langle {\hat \xi} \rangle (t)$ which emerges during the interaction with the IR-pulse, and oscillates with the vibrational frequency $\omega$. For COFCl, $\omega \approx 607 $ B \cite{Tikhonov22}, and the vibrational period is thus $T \approx 0.01 t_0$. 
The elongation becomes maximal if the population is equally distributed between pairs of ro-vibrational states $\ket{0} \ket{\phi_j}$ and $\ket{1} \ket{\phi_j}$, see Eq.(\ref{eq:rot_average_01}). Therefore, the pulse length and field strength are chosen such that after the excitation $|b_{0j}|^2=|b_{1j}|^2=1/2$ for $j=1_{10},0$ (red dotted lines in panel (a)) and all other coefficients are zero. This results in a chiral vibrational wavepacket with maximal elongation
$\langle \hat{\xi} \rangle_{max} / \xi_0 = 1/2$ with $\xi_0=\sqrt{\hbar/(2m\omega)}$ as shown in Fig.~\ref{fig:population_MW_IR_long} (c). The phase of the oscillation depends on the relative phase between the electric fields. The excitation process thus requires a stable constant phase between the microwave and IR pulses. 

\begin{figure}[t]
	\centering
	\includegraphics[width=8cm]{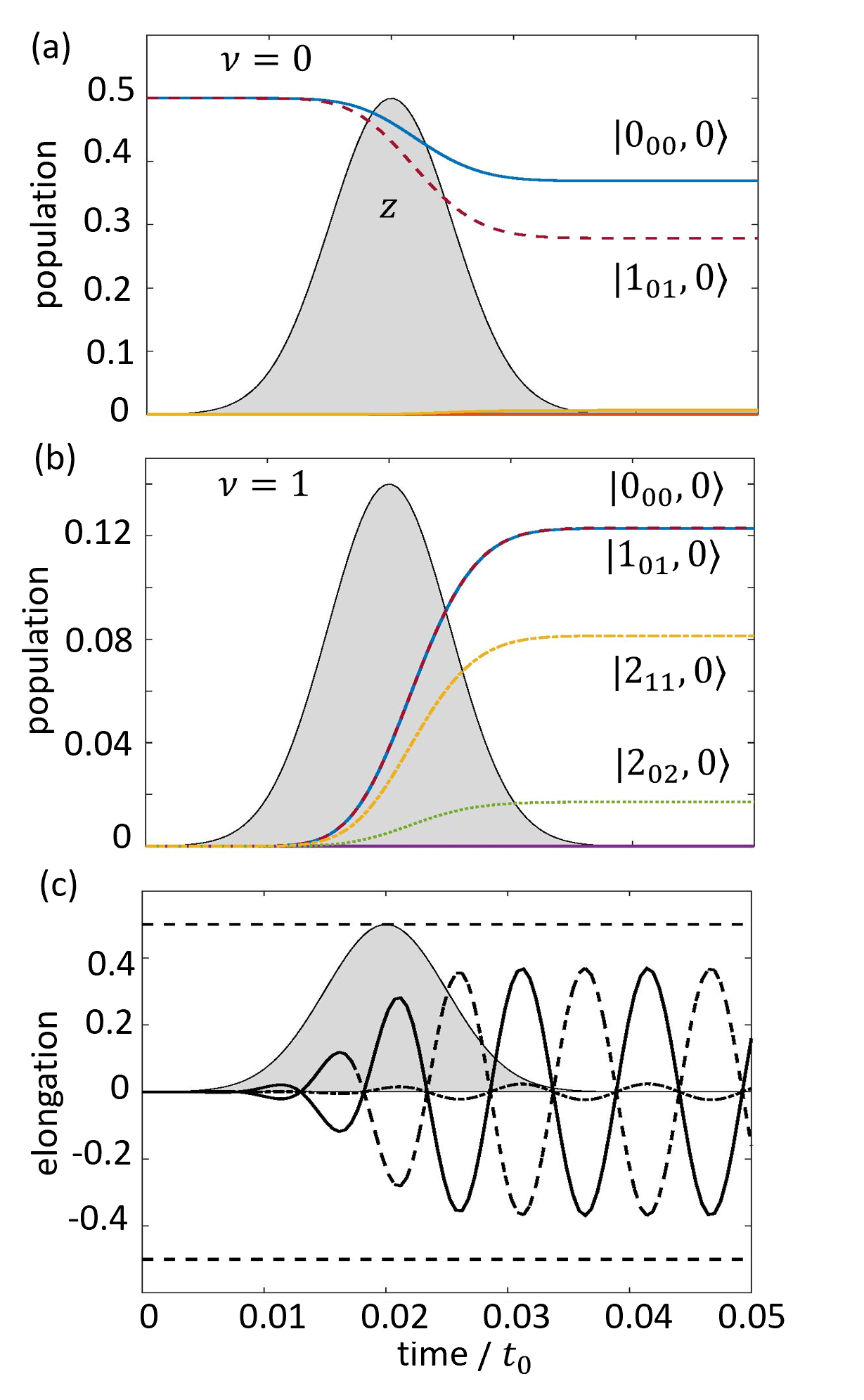}
	\caption{(a,b) Population dynamics after excitation with a short $z$-polarized IR-pulse for the ground (a) and excited (b) vibrational states. The solid blue and dashed red lines correspond to the rotational states $\ket{0_{00},0}$ and $\ket{1_{10},0}$, respectively, the dash-dotted yellow and dotted green lines correspond to $\ket{2_{11},0}$ and $\ket{2_{02},0}$. (c) Elongation $\langle \hat{\xi} \rangle (t) / \xi_0$ due to the excitation with the IR-pulse. The solid blue (dashed green, dash-dotted red) lines correspond to $\phi=\pi/2$ ($-\pi/2$, $0$). The envelope of the IR-pulse is indicated by the gray shapes, the maximal amplitude is ${\cal E}_{IR}=7.5\times 10^7$ V/m.}
	\label{fig:elongation_MW_IR_short}
\end{figure}

To induce a chiral wavepacket, it is, however, not necessary to selectively excite a single ro-vibrational state. As shown in Fig.~\ref{fig:elongation_MW_IR_short}, a chiral wavepacket can also be excited by a short, broadband IR-pulse.
As before, we assume that the rotational wavefunction (\ref{eq:psi_MW}) has been created by microwave excitation. According to the ro-vibrational selection rules, the IR-pulse drives transitions between $\ket{0}\ket{0_{00},0}$ and $\ket{1}\ket{1_{10},0}$ as well as between  $\ket{0} \ket{1_{10},0}$ and $\ket{1} \ket{0_{00},0}$ (see Fig.~\ref{fig:elongation_MW_IR_short} panels (a) and (b)). Choosing the pulse duration and intensity of the IR-pulse such that there is population in all four states after the end of pulse leads to non-vanishing terms in 
Eq.~(\ref{eq:rot_average_01}). The corresponding expectation value $\langle \xi \rangle /\xi_0$ is shown in Fig.~\ref{fig:elongation_MW_IR_short}(c). The average elongation of the out-of-plane coordinate depends on the dynamical phase $\phi$. For $\phi=0$, the average elongation is close to zero, since the contributions to $\langle \xi \rangle$ from the rotational states $\ket{0_{0,0},0}$ and $\ket{1_{1,0},0}$ cancel each other. For  $\phi=\pm \pi/2$, the contributions from the the rotational states $\ket{0_{0,0},0}$ and $\ket{1_{1,0},0}$ add constructively, and $\langle \xi \rangle$ changes its phase if one switches from $\phi=\pi/2$ to $\phi=-\pi/2$. Thus the relative phases between the microwave and IR-pulses must be constant during the interaction. 

It should be noted that the IR-pulse also drives transitions between the initial state $\ket{0} \ket{1_{10},0}$ and the ro-vibrational states $\ket{1} \ket{2_{11},0}$ and $\ket{1} \ket{2_{02},0}$ (see yellow and green lines in panel (b) of Fig.~\ref{fig:population_MW_IR_long}). Since the corresponding states $\ket{0}\ket{2_{11},0}$ and $\ket{0}\ket{2_{02},0}$ are not populated, these states do not contribute to the average elongation of the out-of-plane coordinate, and, as a result, the maximal elongation of the chiral wavepacket is smaller than $1/2$. Further optimization of the shape of the IR-pulse can be used to minimize the population transfer to the $J=2$ states and thus maximize the amplitude of the elongation.

We have numerically demonstrated the excitation of a chiral vibrational wavepacket with two microwave pulses, polarized in $x$- and $y$-direction and one  $z$-polarized IR pulse for a molecule that is initially in its ground rotational and vibrational state. In an experiment, the initial condition will typically contain a (thermal) distribution of rotational states. In this case, the pulse sequences shown here will induce some chiral signal, but it might be small depending on the initial rotational temperature. However, the controllability analysis \ref{subsec:control_MR_IR} shows that it is always possible to induce a chiral wavepacket with maximal elongation. In this case, three microwave fields polarized in $x$-, $y$-, and $z$-direction might be necessary to induce maximal elongation as well as more complicated pulse shapes of the microwave and IR pulses. 

Another experimental challenge for the creation of a chiral wavepacket with microwave and IR pulses is that the phase between the microwave fields and the IR fields must be stable, which can be realized e.g. by using a frequency comb. To avoid phase-locking between microwave and infrared pulses, a chiral wavepacket can also be excited without involving microwave pulses. In Section \ref{sec:example2}, we demonstrate how to create a chiral vibrational wavepacket with a static electric field and a sequence of IR-pulses.

\section{Excitation with IR-pulses in the presence of a static electric field}\label{sec:example2}

In this example, we consider planar
molecules evolving under the influence of the time-dependent Schr\"odinger equation
\begin{equation}
i \hbar \frac{\partial}{\partial t} \ket{\psi(t)} = \left ( H_0 + \epsilon H_{stat} + H_{int}(t) \right ) \ket{\psi(t)},
\label{eq:TDSE_Hstat}
\end{equation}
where
\begin{equation}
 H_{stat} = - {\cal E}_0 {\bm \mu}^{(00)} \cdot {\cal R}(\gamma_R) \cdot {\bm e}_{z} 
\end{equation}
describes the interaction with a static electric field in $z$-direction. Here, the field strength is denoted by ${\cal E}= \epsilon {\cal E}_0$, where ${\cal E}_0$ is a unit field strength and $\epsilon$ a small dimensionless number indicating that the interaction with the static field is small compared to the molecular Hamiltonian $H_0$. The transformation between the space-fixed and molecule-fixed coordinate systems is given by Eq.~(\ref{mu_projection_00}). 
The interaction with the IR pulses can be written as  
\begin{equation}
H_{int}(t) = \sum_i u_i(t) H_i^{IR}, 
\end{equation}
with 
\begin{widetext}
\begin{equation}
  H_i^{IR} = - {\cal E}_i 
  \sum_{j,j'} \ket{0} \ket{\phi_{j'}} \bra{\phi_{j'}} {\bm \mu}^{(01)} \cdot {\cal R}(\gamma_R) \cdot {\bm e}_i  \ket{\phi_{j}} \bra{\phi_{j}} \bra{1} 
     + c.c.,
\end{equation}
\end{widetext}
where  ${\bf E} = {\bf e}_i {\cal E}_i u_i(t)$ is the electric field of the IR-pulse with polarization ${\bf e}_i$, amplitude ${\cal E}_i$ and time-dependence $u_i(t) = s_i(t) \cos(\omega_i t + \phi_i)$.
In Section~\ref{subsec:cond_static_IR}, we show that the Schr\"odinger equation (\ref{eq:TDSE_Hstat}) fulfills the conditions necessary to induce a chiral vibrational wavepacket, as discussed in Section~\ref{subsec:examples}. In Sec.~\ref{subsec:control_static_IR} we analyze the controllability of the Schr\"odinger equation (\ref{eq:TDSE_Hstat}), and in Sec.~\ref{subsec:sim_static_IR}, we numerically demonstrate the excitation of a chiral wavepacket with a sequence of three IR pulses in the presence of a static electric field. 

\subsection{Conditions for creating a chiral wavepacket}\label{subsec:cond_static_IR}
As demonstrated in Section \ref{subsec:examples} (c), a vibrational wavepacket can be excited by three IR pulses with $x$-, $y$-, and $z$-polarization under the condition that the excitation path contains transition matrix elements with $\bra{1} H_{int,\alpha} \ket{0} \neq 0$ for $\alpha=a,b$ and $c$, respectively. 
Since the transition dipole moment for the out-of-plane vibration of a planar molecule is perpendicular to the molecular plane, only
the component $\mu_c^{(01)}$ is non-zero. Thus, IR-pulses only induce transition matrix elements of the form
$\bra{1} H_{int,c} \ket{0}$ and excitation of a chiral wavepacket with only IR pulses is 
not possible without additional external fields. However, in the presence of a static field, additional ro-vibrational transitions occur, which are forbidden under field-free conditions. In order to determine the non-vanishing transition matrix elements in the presence of a static electric field, we consider the interaction with the static field as a perturbation and describe the field-dressed rotational states $\ket{\phi_j^{fd}}$, i.e., the eigenstates of $H_0+ \epsilon H_{stat}$. {\color{black}Using the selection rules associated with $H_{stat}$, one obtains by first-order perturbation theory that},
\begin{equation}\label{eq:eigenstates-fd}
\ket{\phi_j^{fd}} = \ket{\phi_j} +  \epsilon \ket{\phi_j'}+O(\epsilon)
\end{equation}  
with
\begin{equation}
\ket{\phi_j'} = \sum_{k\neq j} \frac{\bra{\phi_j} H_{stat} \ket{\phi_k}}{E_j^{rot} - E_k^{rot}} \ket{\phi_k}, 
\label{eq:wf_perturbation}
\end{equation}
where only those values of $k$ are considered for which $E_j^{rot}\ne E_k^{rot}$. 
Note that Eqs. (33) and (34) hold even if the eigenvalue corresponding to $|\phi_j\rangle$ is degenerate: indeed, the total Hilbert space splits into the direct sum of $H_{stat}$-invariant subspaces (each of them identified by the quantum number $M$), to which the eigenstates $|\phi_j\rangle$ belong, and in each of these subspaces $H_0$ has a simple spectrum.
The transition matrix elements between the field-dressed rotational states in first-order perturbation theory then read
\begin{eqnarray}\label{eq:couplings}
&&\bra{\phi_k^{fd}} \bra{1} H_i^{IR} \ket{0} \ket{\phi_j^{fd}} \nonumber  =
\bra{\phi_k^{rot}} \bra{1} H_i^{IR} \ket{0} \ket{\phi_j^{rot}}  \nonumber \\
&+& \epsilon \big( \bra{\phi_k'} \bra{1} H_i^{IR} \ket{0} \ket{\phi_j} +
 \bra{\phi_k} \bra{1} H_i^{IR} \ket{0} \ket{\phi_j'}\big) \nonumber \\ &+& O(\epsilon).
\end{eqnarray}
Since the permanent dipole moment of COFCl has components $\mu^{(00)}_a\neq 0$ and  $\mu^{(00)}_b \neq 0$, the interaction with the static field can be decomposed into
$ H_{stat} = H_{stat,a} + H_{stat,b}$, and $H_i^{IR} = H_{i,c}^{IR}$.
Inserting Eq.(\ref{eq:wf_perturbation}) we obtain
\begin{widetext}
\begin{eqnarray}
    \bra{\phi_k} \bra{1} H_{i}^{IR} \ket{0} \ket{\phi_j'} &=& 
    \sum_{l\neq j} \frac{1}{E_j^{rot}-E_l^{rot}} 
    \bra{\phi_k} \bra{1} H_{i,c}^{IR} \ket{0} \ket{\phi_l}
    \bra{\phi_l} \bra{0} H_{stat,a} \ket{0} \ket{\phi_j} \nonumber \\
    && + \sum_{l\neq j} \frac{1}{E_j^{rot}-E_l^{rot}} 
    \bra{\phi_k} \bra{1} H_{i,c}^{IR} \ket{0} \ket{\phi_l}
    \bra{\phi_l} \bra{0} H_{stat,b} \ket{0} \ket{\phi_j}.
    \label{eq:transitions_first_order}
\end{eqnarray}
\end{widetext}
The direct products of the irreducible representations of $D_2$ are $B_c\times B_a =B_b$ and $B_c\times B_b =B_a$, cf. Table~\ref{D2}. Therefore, transitions between the field-dressed states driven by IR pulses are governed in first-order perturbation by matrix elements that transform according to $B_a$ and $B_b$ while transition matrix elements between the unperturbed rotational states transform according to $B_c$. Thus, the conditions for creating a vibrational wavepacket with non-vanishing elongation $\langle \hat{\xi} \rangle$ are fulfilled if the molecules interact with three IR pulses  polarized orthogonal to each other in the presence of a static electric field.

\subsection{Controllability analysis}\label{subsec:control_static_IR}
\begin{figure}[t]
	\centering
	\includegraphics[width=8.5cm]{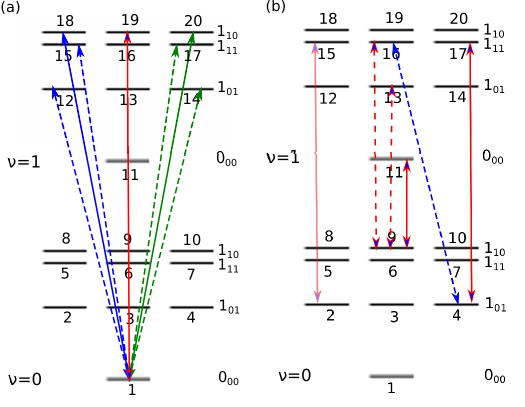}
	\caption{Graph with ro-vibrational states $\ket{J_{k_a,K_c},M} \ket{\nu}$, $\nu=0,1$ as nodes and zero-order (solid lines) and first-order (dashed lines) transition matrix elements as edges. The red, green and blue arrows correspond to transitions induced by control fields with $i=z,\sigma_+$ and $\sigma_-$, respectively. For convenience the nodes are numbered from $1$ to $20$. Panel (a) shows all non-vanishing transitions connecting node $1$. Panel (b) shows an additional set of non-vanishing transitions that is necessary to connect node $1$ to all nodes $\geq 11$.}
	\label{fig:IR_static_graph}
\end{figure}

In order to assure that a maximal chiral signal can be obtained from an arbitrary initial condition, we also study the controllability of the Schr\"odinger equation (\ref{eq:TDSE_Hstat}) using first-order perturbation theory in combination with graphical methods. We therefore introduce a graph  ${\cal G}$ whose nodes are the eigenstates $|\phi_j \rangle|\nu\rangle$ (see horizontal lines in Fig.~\ref{fig:IR_static_graph}). The edges of the graph are defined by the transition matrix elements between the field-dressed eigenstates, Eq.(\ref{eq:couplings}), i.e. by the zero-order couplings $\bra{\phi_k} \bra{1} H_i^{IR} \ket{0} \ket{\phi_j}$ and the first-order couplings $\bra{\phi_k'} \bra{1} H_i^{IR} \ket{0} \ket{\phi_j}+\bra{\phi_k} \bra{1} H_i^{IR} \ket{0} \ket{\phi_j'}$. To analyze the controllability, it is convenient to choose the polarizations $i=\sigma_+,\sigma_-$, where $\sigma_+=x+iy$ and $\sigma_-=x-iy$ instead of $i=x,y$. The Schr\"odinger equation (\ref{eq:TDSE_Hstat}) is 
controllable if the graph ${\cal G}$ has a sub-graph containing all nodes and only {\color{black} uncoupled or} decoupled transitions. It should be noted that if the system is controllable with control fields with $i=z,\sigma_+,\sigma_-$ it is controllable as well with a set of fields with linear polarizations $i=x,y,z$. 

A way to prove controllability is to draw all edges, i.e. all non-zero zero- and first-order transition matrix elements and analyze if the graph ${\cal G}$  is connected by uncoupled or decoupled transitions. It is, however, more convenient to start with a smaller sub-graph and subsequently add as many edges as are necessary to prove that the graph is connected. Therefore, we first analyze all transitions with the ground state  $\ket{0} \ket{\phi_0} =  \ket{0} \ket{0_{0,0},0}$, as indicated in Fig.~\ref{fig:IR_static_graph}.
To simplify the notation, we number the nodes of the graph, starting from $1$ for the ground state $\ket{0} \ket{0_{0,0},0}$ to $20$ for the state $\ket{1} \ket{1_{0,1},1}$, see Fig.~\ref{fig:IR_static_graph}. A $z$-polarized IR pulse induces the zero-order transition $(1\leftrightarrow 19)$, indicated by the solid red line in panel (a). Fields with $i=\sigma_+$ induce transitions with $\Delta M = +1$, i.e. the zero-order transition $(1\leftrightarrow 20)$ and the first-order transitions $(1\leftrightarrow 14)$ and $(1\leftrightarrow 17)$, which are determined by Eq.~(\ref{eq:transitions_first_order}).  Likewise fields with $i=\sigma_-$ induce transitions with $\Delta M = -1$, i.e. the transitions $(1\leftrightarrow 18)$, $(1\leftrightarrow 15)$ and $(1\leftrightarrow 12)$. All those transitions are decoupled since those transitions corresponding to the same control Hamiltonian (same polarization) have different energy gaps. 

Next, we show that the node $1$ is also connected with the nodes $11$, $13$ and $16$. Note that there are no zero- or first-order transitions matrix elements that directly connect vortex $1$ with any of these states. We add the edges $(9\leftrightarrow 11)$, $(9\leftrightarrow 13)$, $(9\leftrightarrow 16)$, $(2\leftrightarrow 15)$, $(4\leftrightarrow 17)$ and $(4\leftrightarrow 16)$, which are shown in Fig.~\ref{fig:IR_static_graph}(b) to the sub-graph. 
All edges represented by red arrows are induced by a $z$-polarized control field, with $(9\leftrightarrow 11)$, $(2\leftrightarrow 15)$ and $(4\leftrightarrow 17)$ being a zero-order transition while $(9\leftrightarrow 13)$ and $(9\leftrightarrow 16)$ are first-order transition, as it can be verified with help of Eq.~(\ref{eq:transitions_first_order}). The zero-order transition $(4\leftrightarrow 16)$ is induced by a control field with polarization $i=\sigma_-$. Since the sub-graph contains the edge $(1\leftrightarrow 17)$, see Fig.~\ref{fig:IR_static_graph} (a), the additional edges couple vortex $1$ with $16$, $13$ and $11$. Note that the transitions $(2\leftrightarrow 15)$ and $(4\leftrightarrow 17)$ are induced by the same control field ($i=z$) and have the same energy gap. They are thus coupled. By taking the graphical commutator between the transition $(1\leftrightarrow 17)$ and $(4\leftrightarrow 17)$, one can, however decouple the transitions $(2\leftrightarrow 15)$ and $(4\leftrightarrow 17)$ \cite{Gago_2023}. The sub-graph consisting of all edges shown in Fig.~\ref{fig:IR_static_graph} (a) and (b) thus contains only decoupled transitions. Node $1$ is thus connected to all nodes in the upper part of the graph, i.e. for all states in the excited vibrational states $\ket{\nu=1}$. 
 Finally, we show that node $11$ is connected with all nodes smaller or equal than $10$: it suffices to do it symmetrically, that is, substitute $1$ with $11$ and any state $n\geq 11$ with $n-10$. The sub-graph we have described satisfies the desired properties: it is a connected graph consisting of only uncoupled or decoupled transitions. This proves, that the corresponding Schr\"odinger equation (\ref{eq:TDSE_Hstat}) is 
 controllable. Controllability of Eq.~(\ref{eq:TDSE_Hstat}) implies that with a combination of a static field and a set of $x,y$ and $z$ or $\sigma_+$, $\sigma_-$ and $z$ polarized IR pulses, a chiral wavepacket with maximal elongation can be created.

 The interaction between the molecule and a static field is typically much smaller than the rotational energies, that is, {\color{black} we consider the regime} $\epsilon \ll 1$. We therefore treated the resulting transition matrix elements in first-order perturbation theory. However, since eigenvalues and eigenvectors of $H_0+\epsilon H_{stat}$ are analytic functions of the parameter $\epsilon$, and analytic functions have at worst isolated zeroes (when they are not identically zero), the controllability result holds for almost every value of the parameter ${\epsilon}$. 

\subsection{Chiral ro-vibrational dynamics in a static field} \label{subsec:sim_static_IR}
\begin{figure}[t]
	\centering
    \includegraphics[width=8cm]{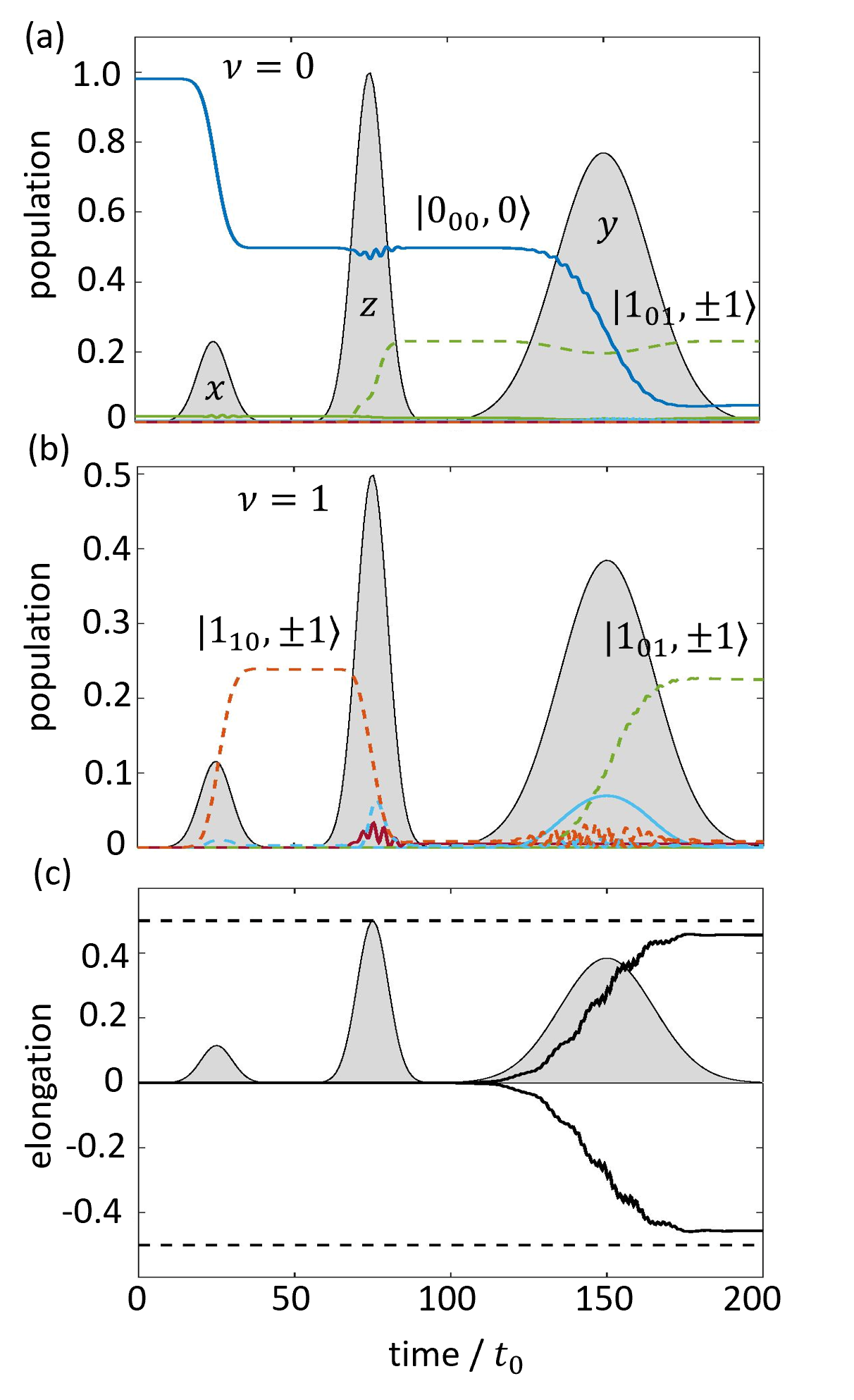}
	\caption{Population dynamics for excitation with a sequence of three IR-pulses in the presence of a static electric field. Panels (a) and (b) show the population in the ground and excited vibrational states. The blue, green and red lines correspond to the rotational states $\ket{0_{00},0}$, $\ket{1_{10},\pm 1}$ and $\ket{1_{01},\pm 1}$, respectively. Panel (c) shows the envelope of the elongation $\langle \xi \rangle(t)/\xi_0$. The dashed lines indicate the maximal elongation $\langle \xi \rangle/\xi_0 = \pm 1/2$. In all panels, the envelopes of the pulses are indicated by the gray shapes, the maximal amplitude is ${\cal E}_i= 1.5 \times 10^5$, $6.5 \times 10^5$ and $5 \times 10^5$ V/m for $i=1,2,3$, respectively. The polarization of the pulses is denotes by $x$, $y$, and $z$. Note that the population dynamics is shown in the field-free basis.}
	\label{fig:pop_static_IR}
\end{figure}

\begin{center}
  \begin{figure}[t]
	\includegraphics[width=16cm]{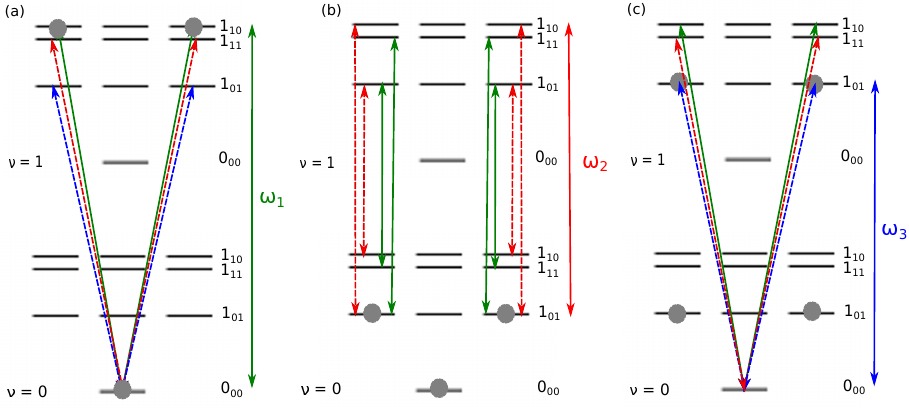}
	\caption{Transition matrix elements and populated states for the excitation process shown in Fig.~\ref{fig:pop_static_IR}. Panels (a), (b), and (c) show the transition matrix elements for the excitation with the IR pulses 1, 2 and 3 of  Fig.~\ref{fig:pop_static_IR}. The solid arrows indicate transition matrix elements between field-free rotational states, dashed arrows correspond to transitions allowed only in the presence of a static electric field. The colors (green, blue, red) indicate the type of transitions ($a$-, $b$- and $c$-type). The gray circles show which states are populated at the end of the first (a), second (b) and third (c) pulse.}
	\label{fig:IR_static_scheme}
\end{figure}
\end{center}

In order to simulate the ro-vibrational dynamics in the presence of a static electric field, we first numerically determine the field-dressed eigenstates $\ket{\phi_j^{fd}} \ket{\nu}$, i.e. the eigenstates of $H_0 + \epsilon H_{stat}$ for $\epsilon=0.3$, which corresponds to a field strength of ${\cal E}=10^6$ V/m.
Similar to Section~\ref{subsec:dynamics_MW_IR}, we assume that the molecule is initially in the field-dressed ro-vibrational ground state $\ket{\psi(0)} = \ket{0} \ket{\phi_0^{fd}}$ and simulate the population dynamics during the interaction with three orthogonally prolarized IR pulses by numerically integrating the
Schr\"odinger equation (\ref{eq:TDSE_Hstat}). Panels (a) and (b) of
Fig.~\ref{fig:pop_static_IR} display the population in the ground and excited vibrational states, respectively. 
The corresponding excitation mechanism is sketched in Fig.~\ref{fig:IR_static_scheme}. 
The first pulse is $x$-polarized and has a Gaussian shape with central frequency $\omega_1=\omega+(E^{rot}_{1_{10}}-E^{rot}_{0_{00}})/\hbar$. The allowed transitions are indicated by the solid and dashed arrows in Fig.~\ref{fig:IR_static_scheme} (a). The solid green arrow represents the transitions matrix elements between the field-free rotational states and the dashed arrows show the much weaker transitions resulting from couplings between rotational states in the presence of the static field. The strength and width of the pulse are chosen such that only the transition resonant to $\omega_1$ is driven and 50 \% of the population remains in the ground ro-vibrational state (blue line in Fig.~\ref{fig:pop_static_IR}) and 25 \% is transferred to the states $\ket{1}\ket{1_{10},\pm 1}$ each. Note that this is a pure c-type transition. To fulfill the conditions for the creation of a chiral wavepacket, the other two pulses have to drive a- and b-type transitions, which arise only due to the couplings induced by the static field and are thus much weaker. The integrated field intensity of these pulses is therefore much larger (see gray shapes in Fig.~\ref{fig:pop_static_IR}). The second pulse is $z$-polarized and has a central frequency $\omega_2=\omega+(E^{rot}_{1_{10}}-E^{rot}_{1_{01}})/\hbar$ and drives the transitions from $\ket{1}\ket{1_{10},\pm 1}$ to $\ket{0}\ket{1_{01},\pm 1}$, which is a b-type transition (see dashed red arrow in panel (b) of Fig.~\ref{fig:IR_static_scheme}). Finally, the third, $y$-polarized pulse transfers the remaining ground state population to $\ket{1}\ket{1_{01},\pm 1}$ by an a-type transition (dashed blue arrows in panel (c)) and thus creates the chiral wavepacket.
The envelope of the elongation $\langle \hat{\xi} \rangle (t) / \xi_0$ is shown in Fig.~\ref{fig:pop_static_IR} (c). The elongation oscillates with frequency $\omega$ (With $T = 2 \pi/\omega \approx 0.01 t_0$ the oscillation is too fast to be resolved on the time-scale shown in Fig.~\ref{fig:pop_static_IR}). Since this pulse sequence transfers the complete population to the two pairs of states  $\ket{0}\ket{1_{01},\pm 1}$ and $\ket{1}\ket{1_{01},\pm 1}$, the elongation reaches its maximal value $\langle \hat{\xi} \rangle/\xi_0 = \pm 1/2$. Note that in the third step, the transition between $\ket{0}\ket{0_{00},0}$ and $\ket{1}\ket{1_{01},\pm 1}$ can also be induced by a second $x$-polarized pulse. 
However, in that case the vibrational wavepacket created by the pair of states $\ket{0}\ket{1_{01},+1}$ and $\ket{1}\ket{1_{01},+1}$
oscillates out of phase with the wavepacket consisting of the states $\ket{0}\ket{1_{01},-1}$ and $\ket{1}\ket{1_{01},-1}$ and thus the net elongation becomes zero. 

We have thus demonstrated numerically that a chiral wavepacket can be excited with a combination of three IR-pulses polarized orthogonal to each other if a static electric field induces transitions between ro-vibrational states which are forbidden under field-free conditions. In this case, the relative phases between the three IR pulses have to be constant, which is easier to realize experimentally than phase-locking between microwave and IR pulses.

\section{Conclusions} \label{sec:conclusions}
With the help of symmetry considerations and controllability analysis we have derived requirements for electromagnetic fields interacting in electric dipole approximation with randomly oriented molecules that allow for the excitation of coherent vibrational motion. In particular we have determined conditions for the creation of chiral vibrational wavepackts in achiral molecules. In order to measure the chirality of the excited molecule, i.e. to measure the time-dependent elongation along the out-of-plane mode, all excitation processes described here could be combined with vibrationally resolved photoelectron circular dichroism, as proposed in \cite{Tikhonov22}. 

The conditions for creating chiral vibrational wavepackets have also been derived by taking a classical rotational average of an ensemble of randomly oriented molecules, where the rotation is assumed to be frozen during the excitation process \cite{Tikhonov22}. The quantum mechanical derivation shown here extends this treatment to cases where molecules rotate during the excitation process. This allows us to compare conditions for creating (temporal) chirality in regimes where molecular rotations behave classically and quantum mechanically. 
The excitation schemes proposed here rely on coherent excitation of individual rotational or ro-vibrational states, i.e. require interaction times long enough to resolve the rotational spectrum of molecules. Such transitions can only be described by a quantum mechanical treatment of molecular rotation. 

While the excitation scheme proposed in \cite{Tikhonov22} requires Raman excitation to an excited electronic state, exciting individual rotational states for the creation of a chiral vibrational wavepacket occurs only in the electronic ground state of the COFCl molecules. This has the advantage that the dipole and transition dipole moments within the electronic ground states are larger than electronic transition dipole moments. 

As proof of principle, we have simulated the excitation of chiral wavepackets assuming that the molecules are initially in their ground rotational and vibrational state. A realistic description of experimental conditions is a initial thermal distribution of rotational states, while the molecules are predominantly in their vibrational ground state. With help of controllability analysis, we have demonstrated that also for such initial condition a chiral signal with maximal amplitude can be achieved with the proposed excitation mechanisms. However, in this case the microwave and IR-pulses are expected to have more complicated shape and can be obtained for example with the help of optimal control theory.

For the controllability analyis we have extended graph-theoretical methods derived for quantum asymmetric tops \cite{Leibscher22,Pozzoli21} to ro-vibrational systems. Perturbation theory for controllability in the presence of a static field is usually considered 
to lift spectral degeneracies, see e.g. \cite{Chambrion22} for such an analysis in rotating symmetric top molecules. Here, we pursue a different approach, where perturbation
theory for controllability in the presence of a static field is
considered to create additional couplings in the matrix elements, which graphically represent additional edges of the quantum spectral graph.

\onecolumngrid
\twocolumngrid

\begin{acknowledgments}
  The ideas for this study are based on inspiring discussions with D. Tikhonov and M. Schnell for which we are very thankful.
  We gratefully acknowledge financial support from the Deutsche Forschungsgemeinschaft through CRC 1319 ELCH and through the joint ANR-DFG CoRoMo project 505622963 / KO 2301/15-1, ANR-22-CE92-0077-01. 
   MS and UB also thank the ANR project Quaco ANR-17-CE40-0007-01. EP thanks the project CONSTAT, supported by the Conseil Régional de Bourgogne Franche-Comté and the European Union through the PO FEDER Bourgogne 2014/2020 programs, the EIPHI Graduate School (ANR-17-EURE-0002), the STARS Consolidator Grant 2021 “NewSRG” of the University of Padova, and the PNRR MUR project PE0000023-NQSTI.
\end{acknowledgments}

\bibliographystyle{apsrev4-1}

\begin{thebibliography}{49}%
	\makeatletter
	\providecommand \@ifxundefined [1]{%
		\@ifx{#1\undefined}
	}%
	\providecommand \@ifnum [1]{%
		\ifnum #1\expandafter \@firstoftwo
		\else \expandafter \@secondoftwo
		\fi
	}%
	\providecommand \@ifx [1]{%
		\ifx #1\expandafter \@firstoftwo
		\else \expandafter \@secondoftwo
		\fi
	}%
	\providecommand \natexlab [1]{#1}%
	\providecommand \enquote  [1]{``#1''}%
	\providecommand \bibnamefont  [1]{#1}%
	\providecommand \bibfnamefont [1]{#1}%
	\providecommand \citenamefont [1]{#1}%
	\providecommand \href@noop [0]{\@secondoftwo}%
	\providecommand \href [0]{\begingroup \@sanitize@url \@href}%
	\providecommand \@href[1]{\@@startlink{#1}\@@href}%
	\providecommand \@@href[1]{\endgroup#1\@@endlink}%
	\providecommand \@sanitize@url [0]{\catcode `\\12\catcode `\$12\catcode
		`\&12\catcode `\#12\catcode `\^12\catcode `\_12\catcode `\%12\relax}%
	\providecommand \@@startlink[1]{}%
	\providecommand \@@endlink[0]{}%
	\providecommand \url  [0]{\begingroup\@sanitize@url \@url }%
	\providecommand \@url [1]{\endgroup\@href {#1}{\urlprefix }}%
	\providecommand \urlprefix  [0]{URL }%
	\providecommand \Eprint [0]{\href }%
	\providecommand \doibase [0]{http://dx.doi.org/}%
	\providecommand \selectlanguage [0]{\@gobble}%
	\providecommand \bibinfo  [0]{\@secondoftwo}%
	\providecommand \bibfield  [0]{\@secondoftwo}%
	\providecommand \translation [1]{[#1]}%
	\providecommand \BibitemOpen [0]{}%
	\providecommand \bibitemStop [0]{}%
	\providecommand \bibitemNoStop [0]{.\EOS\space}%
	\providecommand \EOS [0]{\spacefactor3000\relax}%
	\providecommand \BibitemShut  [1]{\csname bibitem#1\endcsname}%
	\let\auto@bib@innerbib\@empty
	\bibitem [{\citenamefont {Lux}\ \emph {et~al.}(2012)\citenamefont {Lux},
		\citenamefont {Wollenhaupt}, \citenamefont {Bolze}, \citenamefont {Liang},
		\citenamefont {K\"ohler}, \citenamefont {Sarpe},\ and\ \citenamefont
		{Baumert}}]{LuxAngewandte12}%
	\BibitemOpen
	\bibfield  {author} {\bibinfo {author} {\bibfnamefont {C.}~\bibnamefont
			{Lux}}, \bibinfo {author} {\bibfnamefont {M.}~\bibnamefont {Wollenhaupt}},
		\bibinfo {author} {\bibfnamefont {T.}~\bibnamefont {Bolze}}, \bibinfo
		{author} {\bibfnamefont {Q.}~\bibnamefont {Liang}}, \bibinfo {author}
		{\bibfnamefont {J.}~\bibnamefont {K\"ohler}}, \bibinfo {author}
		{\bibfnamefont {C.}~\bibnamefont {Sarpe}}, \ and\ \bibinfo {author}
		{\bibfnamefont {T.}~\bibnamefont {Baumert}},\ }\href {\doibase
		10.1002/anie.201109035} {\bibfield  {journal} {\bibinfo  {journal} {Angew.
				Chem. Int. Ed.}\ }\textbf {\bibinfo {volume} {51}},\ \bibinfo {pages} {5001}
		(\bibinfo {year} {2012})}\BibitemShut {NoStop}%
	\bibitem [{\citenamefont {Patterson}\ \emph {et~al.}(2013)\citenamefont
		{Patterson}, \citenamefont {Schnell},\ and\ \citenamefont
		{Doyle}}]{PattersonNature13}%
	\BibitemOpen
	\bibfield  {author} {\bibinfo {author} {\bibfnamefont {D.}~\bibnamefont
			{Patterson}}, \bibinfo {author} {\bibfnamefont {M.}~\bibnamefont {Schnell}},
		\ and\ \bibinfo {author} {\bibfnamefont {J.~M.}\ \bibnamefont {Doyle}},\
	}\href {\doibase 10.1038/nature12150} {\bibfield  {journal} {\bibinfo
			{journal} {Nature}\ }\textbf {\bibinfo {volume} {497}},\ \bibinfo {pages}
		{475} (\bibinfo {year} {2013})}\BibitemShut {NoStop}%
	\bibitem [{\citenamefont {Cireasa}\ \emph {et~al.}(2015)\citenamefont
		{Cireasa}, \citenamefont {Boguslavskiy}, \citenamefont {Pons}, \citenamefont
		{Wong}, \citenamefont {Descamps}, \citenamefont {Petit}, \citenamefont {Ruf},
		\citenamefont {Thir\'e}, \citenamefont {Ferr\'e}, \citenamefont {Suarez},
		\citenamefont {Higuet}, \citenamefont {Schmidt}, \citenamefont {Alharbi},
		\citenamefont {L\'egar\'e}, \citenamefont {Blanchet}, \citenamefont {Fabre},
		\citenamefont {Patchkovskii}, \citenamefont {Smirnova}, \citenamefont
		{Mairesse},\ and\ \citenamefont {Bhardwaj}}]{CireasaNatPhys15}%
	\BibitemOpen
	\bibfield  {author} {\bibinfo {author} {\bibfnamefont {R.}~\bibnamefont
			{Cireasa}}, \bibinfo {author} {\bibfnamefont {A.~E.}\ \bibnamefont
			{Boguslavskiy}}, \bibinfo {author} {\bibfnamefont {B.}~\bibnamefont {Pons}},
		\bibinfo {author} {\bibfnamefont {M.~C.~H.}\ \bibnamefont {Wong}}, \bibinfo
		{author} {\bibfnamefont {D.}~\bibnamefont {Descamps}}, \bibinfo {author}
		{\bibfnamefont {S.}~\bibnamefont {Petit}}, \bibinfo {author} {\bibfnamefont
			{H.}~\bibnamefont {Ruf}}, \bibinfo {author} {\bibfnamefont {N.}~\bibnamefont
			{Thir\'e}}, \bibinfo {author} {\bibfnamefont {A.}~\bibnamefont {Ferr\'e}},
		\bibinfo {author} {\bibfnamefont {J.}~\bibnamefont {Suarez}}, \bibinfo
		{author} {\bibfnamefont {J.}~\bibnamefont {Higuet}}, \bibinfo {author}
		{\bibfnamefont {B.~E.}\ \bibnamefont {Schmidt}}, \bibinfo {author}
		{\bibfnamefont {A.~F.}\ \bibnamefont {Alharbi}}, \bibinfo {author}
		{\bibfnamefont {F.}~\bibnamefont {L\'egar\'e}}, \bibinfo {author}
		{\bibfnamefont {V.}~\bibnamefont {Blanchet}}, \bibinfo {author}
		{\bibfnamefont {B.}~\bibnamefont {Fabre}}, \bibinfo {author} {\bibfnamefont
			{S.}~\bibnamefont {Patchkovskii}}, \bibinfo {author} {\bibfnamefont
			{O.}~\bibnamefont {Smirnova}}, \bibinfo {author} {\bibfnamefont
			{Y.}~\bibnamefont {Mairesse}}, \ and\ \bibinfo {author} {\bibfnamefont
			{V.~R.}\ \bibnamefont {Bhardwaj}},\ }\href {\doibase 10.1038/nphys3369}
	{\bibfield  {journal} {\bibinfo  {journal} {Nat. Phys.}\ }\textbf {\bibinfo
			{volume} {11}},\ \bibinfo {pages} {654} (\bibinfo {year} {2015})}\BibitemShut
	{NoStop}%
	\bibitem [{\citenamefont {Shubert}\ \emph {et~al.}(2014)\citenamefont
		{Shubert}, \citenamefont {Schmitz}, \citenamefont {Patterson}, \citenamefont
		{Doyle},\ and\ \citenamefont {Schnell}}]{ShubertAngewandte14}%
	\BibitemOpen
	\bibfield  {author} {\bibinfo {author} {\bibfnamefont {V.~A.}\ \bibnamefont
			{Shubert}}, \bibinfo {author} {\bibfnamefont {D.}~\bibnamefont {Schmitz}},
		\bibinfo {author} {\bibfnamefont {D.}~\bibnamefont {Patterson}}, \bibinfo
		{author} {\bibfnamefont {J.~M.}\ \bibnamefont {Doyle}}, \ and\ \bibinfo
		{author} {\bibfnamefont {M.}~\bibnamefont {Schnell}},\ }\href {\doibase
		10.1002/anie.201306271} {\bibfield  {journal} {\bibinfo  {journal} {Angew.
				Chem. Int. Ed.}\ }\textbf {\bibinfo {volume} {53}},\ \bibinfo {pages} {1152}
		(\bibinfo {year} {2014})}\BibitemShut {NoStop}%
	\bibitem [{\citenamefont {Lobsiger}\ \emph {et~al.}(2015)\citenamefont
		{Lobsiger}, \citenamefont {P{\'e}rez}, \citenamefont {Evangelisti},
		\citenamefont {Lehmann},\ and\ \citenamefont {Pate}}]{LobsigerJPCL15}%
	\BibitemOpen
	\bibfield  {author} {\bibinfo {author} {\bibfnamefont {S.}~\bibnamefont
			{Lobsiger}}, \bibinfo {author} {\bibfnamefont {C.}~\bibnamefont {P{\'e}rez}},
		\bibinfo {author} {\bibfnamefont {L.}~\bibnamefont {Evangelisti}}, \bibinfo
		{author} {\bibfnamefont {K.~K.}\ \bibnamefont {Lehmann}}, \ and\ \bibinfo
		{author} {\bibfnamefont {B.~H.}\ \bibnamefont {Pate}},\ }\href {\doibase
		10.1021/jz502312t} {\bibfield  {journal} {\bibinfo  {journal} {J. Phys. Chem.
				Lett.}\ }\textbf {\bibinfo {volume} {6}},\ \bibinfo {pages} {196} (\bibinfo
		{year} {2015})}\BibitemShut {NoStop}%
	\bibitem [{\citenamefont {Milner}\ \emph {et~al.}(2019)\citenamefont {Milner},
		\citenamefont {Fordyce}, \citenamefont {MacPhail-Bartley}, \citenamefont
		{Wasserman}, \citenamefont {Milner}, \citenamefont {Tutunnikov},\ and\
		\citenamefont {Averbukh}}]{MilnerPRL2019}%
	\BibitemOpen
	\bibfield  {author} {\bibinfo {author} {\bibfnamefont {A.}~\bibnamefont
			{Milner}}, \bibinfo {author} {\bibfnamefont {J.~A.~M.}\ \bibnamefont
			{Fordyce}}, \bibinfo {author} {\bibnamefont {MacPhail-Bartley}}, \bibinfo
		{author} {\bibfnamefont {W.}~\bibnamefont {Wasserman}}, \bibinfo {author}
		{\bibfnamefont {V.}~\bibnamefont {Milner}}, \bibinfo {author} {\bibfnamefont
			{I.}~\bibnamefont {Tutunnikov}}, \ and\ \bibinfo {author} {\bibfnamefont
			{I.~S.}\ \bibnamefont {Averbukh}},\ }\href {\doibase
		10.1103/PhysRevLett.122.223201} {\bibfield  {journal} {\bibinfo  {journal}
			{Phys. Rev. Lett.}\ }\textbf {\bibinfo {volume} {122}},\ \bibinfo {pages}
		{223201} (\bibinfo {year} {2019})}\BibitemShut {NoStop}%
	\bibitem [{\citenamefont {Lee}\ \emph {et~al.}(2022)\citenamefont {Lee},
		\citenamefont {Bischoff}, \citenamefont {Hernandez-Castillo}, \citenamefont
		{Sartakov}, \citenamefont {Meijer},\ and\ \citenamefont
		{Eibenberger-Arias}}]{Lee21}%
	\BibitemOpen
	\bibfield  {author} {\bibinfo {author} {\bibfnamefont {J.}~\bibnamefont
			{Lee}}, \bibinfo {author} {\bibfnamefont {J.}~\bibnamefont {Bischoff}},
		\bibinfo {author} {\bibfnamefont {A.~O.}\ \bibnamefont {Hernandez-Castillo}},
		\bibinfo {author} {\bibfnamefont {B.}~\bibnamefont {Sartakov}}, \bibinfo
		{author} {\bibfnamefont {G.}~\bibnamefont {Meijer}}, \ and\ \bibinfo {author}
		{\bibfnamefont {S.}~\bibnamefont {Eibenberger-Arias}},\ }\href {\doibase
		10.1103/PhysRevLett.128.173001} {\bibfield  {journal} {\bibinfo  {journal}
			{Phys. Rev. Lett}\ }\textbf {\bibinfo {volume} {128}},\ \bibinfo {pages}
		{173001} (\bibinfo {year} {2022})}\BibitemShut {NoStop}%
	\bibitem [{\citenamefont {Singh}\ \emph {et~al.}(2023)\citenamefont {Singh},
		\citenamefont {Bergg\"otz}, \citenamefont {Sun},\ and\ \citenamefont
		{Schnell}}]{Singh_Angw_2023}%
	\BibitemOpen
	\bibfield  {author} {\bibinfo {author} {\bibfnamefont {H.}~\bibnamefont
			{Singh}}, \bibinfo {author} {\bibfnamefont {F.~E.~L.}\ \bibnamefont
			{Bergg\"otz}}, \bibinfo {author} {\bibfnamefont {W.}~\bibnamefont {Sun}}, \
		and\ \bibinfo {author} {\bibfnamefont {M.}~\bibnamefont {Schnell}},\ }\href
	{https://doi.org/10.1002/anie.202219045} {\bibfield  {journal} {\bibinfo
			{journal} {Angew. Chem.Int. Ed.}\ }\textbf {\bibinfo {volume} {62}},\
		\bibinfo {pages} {e202219045} (\bibinfo {year} {2023})}\BibitemShut {NoStop}%
	\bibitem [{\citenamefont {Faccial\`a}\ \emph {et~al.}(2023)\citenamefont
		{Faccial\`a}, \citenamefont {Devetta}, \citenamefont {Beauvarlet},
		\citenamefont {Besley}, \citenamefont {Calegari}, \citenamefont {Callegari},
		\citenamefont {Catone}, \citenamefont {Cinquanta}, \citenamefont {Ciriolo},
		\citenamefont {Colaizzi}, \citenamefont {Coreno}, \citenamefont {Crippa},
		\citenamefont {De~Ninno}, \citenamefont {Di~Fraia}, \citenamefont {Galli},
		\citenamefont {Garcia}, \citenamefont {Mairesse}, \citenamefont {Negro},
		\citenamefont {Plekan}, \citenamefont {Prasannan~Geetha}, \citenamefont
		{Prince}, \citenamefont {Pusala}, \citenamefont {Stagira}, \citenamefont
		{Turchini}, \citenamefont {Ueda}, \citenamefont {You}, \citenamefont {Zema},
		\citenamefont {Blanchet}, \citenamefont {Nahon}, \citenamefont {Powis},\ and\
		\citenamefont {Vozzi}}]{FaccialaPRX23}%
	\BibitemOpen
	\bibfield  {author} {\bibinfo {author} {\bibfnamefont {D.}~\bibnamefont
			{Faccial\`a}}, \bibinfo {author} {\bibfnamefont {M.}~\bibnamefont {Devetta}},
		\bibinfo {author} {\bibfnamefont {S.}~\bibnamefont {Beauvarlet}}, \bibinfo
		{author} {\bibfnamefont {N.}~\bibnamefont {Besley}}, \bibinfo {author}
		{\bibfnamefont {F.}~\bibnamefont {Calegari}}, \bibinfo {author}
		{\bibfnamefont {C.}~\bibnamefont {Callegari}}, \bibinfo {author}
		{\bibfnamefont {D.}~\bibnamefont {Catone}}, \bibinfo {author} {\bibfnamefont
			{E.}~\bibnamefont {Cinquanta}}, \bibinfo {author} {\bibfnamefont {A.~G.}\
			\bibnamefont {Ciriolo}}, \bibinfo {author} {\bibfnamefont {L.}~\bibnamefont
			{Colaizzi}}, \bibinfo {author} {\bibfnamefont {M.}~\bibnamefont {Coreno}},
		\bibinfo {author} {\bibfnamefont {G.}~\bibnamefont {Crippa}}, \bibinfo
		{author} {\bibfnamefont {G.}~\bibnamefont {De~Ninno}}, \bibinfo {author}
		{\bibfnamefont {M.}~\bibnamefont {Di~Fraia}}, \bibinfo {author}
		{\bibfnamefont {M.}~\bibnamefont {Galli}}, \bibinfo {author} {\bibfnamefont
			{G.~A.}\ \bibnamefont {Garcia}}, \bibinfo {author} {\bibfnamefont
			{Y.}~\bibnamefont {Mairesse}}, \bibinfo {author} {\bibfnamefont
			{M.}~\bibnamefont {Negro}}, \bibinfo {author} {\bibfnamefont
			{O.}~\bibnamefont {Plekan}}, \bibinfo {author} {\bibfnamefont
			{P.}~\bibnamefont {Prasannan~Geetha}}, \bibinfo {author} {\bibfnamefont
			{K.~C.}\ \bibnamefont {Prince}}, \bibinfo {author} {\bibfnamefont
			{A.}~\bibnamefont {Pusala}}, \bibinfo {author} {\bibfnamefont
			{S.}~\bibnamefont {Stagira}}, \bibinfo {author} {\bibfnamefont
			{S.}~\bibnamefont {Turchini}}, \bibinfo {author} {\bibfnamefont
			{K.}~\bibnamefont {Ueda}}, \bibinfo {author} {\bibfnamefont {D.}~\bibnamefont
			{You}}, \bibinfo {author} {\bibfnamefont {N.}~\bibnamefont {Zema}}, \bibinfo
		{author} {\bibfnamefont {V.}~\bibnamefont {Blanchet}}, \bibinfo {author}
		{\bibfnamefont {L.}~\bibnamefont {Nahon}}, \bibinfo {author} {\bibfnamefont
			{I.}~\bibnamefont {Powis}}, \ and\ \bibinfo {author} {\bibfnamefont
			{C.}~\bibnamefont {Vozzi}},\ }\href {\doibase 10.1103/PhysRevX.13.011044}
	{\bibfield  {journal} {\bibinfo  {journal} {Phys. Rev. X}\ }\textbf {\bibinfo
			{volume} {13}},\ \bibinfo {pages} {011044} (\bibinfo {year}
		{2023})}\BibitemShut {NoStop}%
	\bibitem [{\citenamefont {Goetz}\ \emph {et~al.}(2017)\citenamefont {Goetz},
		\citenamefont {Isaev}, \citenamefont {Nikoobakht}, \citenamefont {Berger},\
		and\ \citenamefont {Koch}}]{GoetzJCP17}%
	\BibitemOpen
	\bibfield  {author} {\bibinfo {author} {\bibfnamefont {R.~E.}\ \bibnamefont
			{Goetz}}, \bibinfo {author} {\bibfnamefont {T.~A.}\ \bibnamefont {Isaev}},
		\bibinfo {author} {\bibfnamefont {B.}~\bibnamefont {Nikoobakht}}, \bibinfo
		{author} {\bibfnamefont {R.}~\bibnamefont {Berger}}, \ and\ \bibinfo {author}
		{\bibfnamefont {C.~P.}\ \bibnamefont {Koch}},\ }\href {\doibase
		10.1063/1.4973456} {\bibfield  {journal} {\bibinfo  {journal} {J. Chem.
				Phys.}\ }\textbf {\bibinfo {volume} {146}},\ \bibinfo {pages} {024306}
		(\bibinfo {year} {2017})}\BibitemShut {NoStop}%
	\bibitem [{\citenamefont {Demekhin}\ \emph {et~al.}(2018)\citenamefont
		{Demekhin}, \citenamefont {Artemyev}, \citenamefont {Kastner},\ and\
		\citenamefont {Baumert}}]{Demekhin_PRL_2018}%
	\BibitemOpen
	\bibfield  {author} {\bibinfo {author} {\bibfnamefont {P.~V.}\ \bibnamefont
			{Demekhin}}, \bibinfo {author} {\bibfnamefont {A.~N.}\ \bibnamefont
			{Artemyev}}, \bibinfo {author} {\bibfnamefont {A.}~\bibnamefont {Kastner}}, \
		and\ \bibinfo {author} {\bibfnamefont {T.}~\bibnamefont {Baumert}},\ }\href
	{\doibase 10.1103/PhysRevLett.121.253201} {\bibfield  {journal} {\bibinfo
			{journal} {Phys. Rev. Lett.}\ }\textbf {\bibinfo {volume} {121}},\ \bibinfo
		{pages} {253201} (\bibinfo {year} {2018})}\BibitemShut {NoStop}%
	\bibitem [{\citenamefont {Tutunnikov}\ \emph {et~al.}(2018)\citenamefont
		{Tutunnikov}, \citenamefont {Gershnabel}, \citenamefont {Gold},\ and\
		\citenamefont {Averbukh}}]{TutunnikovJPCL18}%
	\BibitemOpen
	\bibfield  {author} {\bibinfo {author} {\bibfnamefont {I.}~\bibnamefont
			{Tutunnikov}}, \bibinfo {author} {\bibfnamefont {E.}~\bibnamefont
			{Gershnabel}}, \bibinfo {author} {\bibfnamefont {S.}~\bibnamefont {Gold}}, \
		and\ \bibinfo {author} {\bibfnamefont {I.~S.}\ \bibnamefont {Averbukh}},\
	}\href {\doibase 10.1021/acs.jpclett.7b03416} {\bibfield  {journal} {\bibinfo
			{journal} {J. Phys. Chem. Lett.}\ }\textbf {\bibinfo {volume} {9}},\
		\bibinfo {pages} {1105} (\bibinfo {year} {2018})}\BibitemShut {NoStop}%
	\bibitem [{\citenamefont {Lehmann}(2018)}]{LehmannJCP18}%
	\BibitemOpen
	\bibfield  {author} {\bibinfo {author} {\bibfnamefont {K.~K.}\ \bibnamefont
			{Lehmann}},\ }\href {\doibase 10.1063/1.5045052} {\bibfield  {journal}
		{\bibinfo  {journal} {J. Chem. Phys.}\ }\textbf {\bibinfo {volume} {149}},\
		\bibinfo {pages} {094201} (\bibinfo {year} {2018})}\BibitemShut {NoStop}%
	\bibitem [{\citenamefont {Leibscher}\ \emph {et~al.}(2019)\citenamefont
		{Leibscher}, \citenamefont {Giesen},\ and\ \citenamefont
		{Koch}}]{Leibscher19}%
	\BibitemOpen
	\bibfield  {author} {\bibinfo {author} {\bibfnamefont {M.}~\bibnamefont
			{Leibscher}}, \bibinfo {author} {\bibfnamefont {T.~F.}\ \bibnamefont
			{Giesen}}, \ and\ \bibinfo {author} {\bibfnamefont {C.~P.}\ \bibnamefont
			{Koch}},\ }\href {\doibase 10.1063/1.5097406} {\bibfield  {journal} {\bibinfo
			{journal} {J. Chem. Phys.}\ }\textbf {\bibinfo {volume} {151}},\ \bibinfo
		{pages} {014302} (\bibinfo {year} {2019})}\BibitemShut {NoStop}%
	\bibitem [{\citenamefont {Ordonez}\ and\ \citenamefont
		{Smirnova}(2019{\natexlab{a}})}]{OrdonezPRA19}%
	\BibitemOpen
	\bibfield  {author} {\bibinfo {author} {\bibfnamefont {A.~F.}\ \bibnamefont
			{Ordonez}}\ and\ \bibinfo {author} {\bibfnamefont {O.}~\bibnamefont
			{Smirnova}},\ }\href {\doibase 10.1103/PhysRevA.99.043416} {\bibfield
		{journal} {\bibinfo  {journal} {Phys. Rev. A}\ }\textbf {\bibinfo {volume}
			{99}},\ \bibinfo {pages} {043416} (\bibinfo {year}
		{2019}{\natexlab{a}})}\BibitemShut {NoStop}%
	\bibitem [{\citenamefont {Goetz}\ \emph {et~al.}(2019)\citenamefont {Goetz},
		\citenamefont {Koch},\ and\ \citenamefont {Greenman}}]{GoetzPRL19}%
	\BibitemOpen
	\bibfield  {author} {\bibinfo {author} {\bibfnamefont {R.~E.}\ \bibnamefont
			{Goetz}}, \bibinfo {author} {\bibfnamefont {C.~P.}\ \bibnamefont {Koch}}, \
		and\ \bibinfo {author} {\bibfnamefont {L.}~\bibnamefont {Greenman}},\ }\href
	{\doibase 10.1103/PhysRevLett.122.013204} {\bibfield  {journal} {\bibinfo
			{journal} {Phys. Rev. Lett.}\ }\textbf {\bibinfo {volume} {122}},\ \bibinfo
		{pages} {013204} (\bibinfo {year} {2019})}\BibitemShut {NoStop}%
	\bibitem [{\citenamefont {Neufeld}\ \emph {et~al.}(2019)\citenamefont
		{Neufeld}, \citenamefont {Ayuso}, \citenamefont {Decleva}, \citenamefont
		{Ivanov}, \citenamefont {Smirnova},\ and\ \citenamefont
		{Cohen}}]{NeufeldPRX19}%
	\BibitemOpen
	\bibfield  {author} {\bibinfo {author} {\bibfnamefont {O.}~\bibnamefont
			{Neufeld}}, \bibinfo {author} {\bibfnamefont {D.}~\bibnamefont {Ayuso}},
		\bibinfo {author} {\bibfnamefont {P.}~\bibnamefont {Decleva}}, \bibinfo
		{author} {\bibfnamefont {M.~Y.}\ \bibnamefont {Ivanov}}, \bibinfo {author}
		{\bibfnamefont {O.}~\bibnamefont {Smirnova}}, \ and\ \bibinfo {author}
		{\bibfnamefont {O.}~\bibnamefont {Cohen}},\ }\href {\doibase
		10.1103/PhysRevX.9.031002} {\bibfield  {journal} {\bibinfo  {journal} {Phys.
				Rev. X}\ }\textbf {\bibinfo {volume} {9}},\ \bibinfo {pages} {031002}
		(\bibinfo {year} {2019})}\BibitemShut {NoStop}%
	\bibitem [{\citenamefont {Vogwell}\ \emph {et~al.}(2023)\citenamefont
		{Vogwell}, \citenamefont {Rego}, \citenamefont {Smirnova},\ and\
		\citenamefont {Ayuso}}]{VogwellSciAdv21}%
	\BibitemOpen
	\bibfield  {author} {\bibinfo {author} {\bibfnamefont {J.}~\bibnamefont
			{Vogwell}}, \bibinfo {author} {\bibfnamefont {L.}~\bibnamefont {Rego}},
		\bibinfo {author} {\bibfnamefont {O.}~\bibnamefont {Smirnova}}, \ and\
		\bibinfo {author} {\bibfnamefont {D.}~\bibnamefont {Ayuso}},\ }\href
	{\doibase 10.1126/sciadv.adj1429} {\bibfield  {journal} {\bibinfo  {journal}
			{Science Advances}\ }\textbf {\bibinfo {volume} {9}},\ \bibinfo {pages}
		{eadj1429} (\bibinfo {year} {2023})}\BibitemShut {NoStop}%
	\bibitem [{\citenamefont {Kastner}\ \emph {et~al.}(2017)\citenamefont
		{Kastner}, \citenamefont {Ring}, \citenamefont {Krueger}, \citenamefont
		{Park}, \citenamefont {Schaefer}, \citenamefont {Senftleben},\ and\
		\citenamefont {Baumert}}]{Kastner_JCP_2017}%
	\BibitemOpen
	\bibfield  {author} {\bibinfo {author} {\bibfnamefont {A.}~\bibnamefont
			{Kastner}}, \bibinfo {author} {\bibfnamefont {T.}~\bibnamefont {Ring}},
		\bibinfo {author} {\bibfnamefont {B.~C.}\ \bibnamefont {Krueger}}, \bibinfo
		{author} {\bibfnamefont {G.~B.}\ \bibnamefont {Park}}, \bibinfo {author}
		{\bibfnamefont {T.}~\bibnamefont {Schaefer}}, \bibinfo {author}
		{\bibfnamefont {A.}~\bibnamefont {Senftleben}}, \ and\ \bibinfo {author}
		{\bibfnamefont {T.}~\bibnamefont {Baumert}},\ }\href
	{https://doi.org/10.1063/1.4982614} {\bibfield  {journal} {\bibinfo
			{journal} {J. Chem. Phys.}\ }\textbf {\bibinfo {volume} {147}},\ \bibinfo
		{pages} {013926} (\bibinfo {year} {2017})}\BibitemShut {NoStop}%
	\bibitem [{\citenamefont {Beaulieu}\ \emph {et~al.}(2018)\citenamefont
		{Beaulieu}, \citenamefont {Comby}, \citenamefont {Descamps}, \citenamefont
		{Fabre}, \citenamefont {Garcia}, \citenamefont {Geneaux}, \citenamefont
		{Harvey}, \citenamefont {Legare}, \citenamefont {Masin}, \citenamefont
		{Nahon}, \citenamefont {Ordonez}, \citenamefont {Petit}, \citenamefont
		{Pons}, \citenamefont {Mairesse}, \citenamefont {Smirnova},\ and\
		\citenamefont {Blanchet}}]{Beaulieu_Nature_2018}%
	\BibitemOpen
	\bibfield  {author} {\bibinfo {author} {\bibfnamefont {S.}~\bibnamefont
			{Beaulieu}}, \bibinfo {author} {\bibfnamefont {A.}~\bibnamefont {Comby}},
		\bibinfo {author} {\bibfnamefont {D.}~\bibnamefont {Descamps}}, \bibinfo
		{author} {\bibfnamefont {B.}~\bibnamefont {Fabre}}, \bibinfo {author}
		{\bibfnamefont {G.~A.}\ \bibnamefont {Garcia}}, \bibinfo {author}
		{\bibfnamefont {R.}~\bibnamefont {Geneaux}}, \bibinfo {author} {\bibfnamefont
			{A.~G.}\ \bibnamefont {Harvey}}, \bibinfo {author} {\bibfnamefont
			{F.}~\bibnamefont {Legare}}, \bibinfo {author} {\bibfnamefont
			{Z.}~\bibnamefont {Masin}}, \bibinfo {author} {\bibfnamefont
			{L.}~\bibnamefont {Nahon}}, \bibinfo {author} {\bibfnamefont {A.~F.}\
			\bibnamefont {Ordonez}}, \bibinfo {author} {\bibfnamefont {S.}~\bibnamefont
			{Petit}}, \bibinfo {author} {\bibfnamefont {B.}~\bibnamefont {Pons}},
		\bibinfo {author} {\bibfnamefont {Y.}~\bibnamefont {Mairesse}}, \bibinfo
		{author} {\bibfnamefont {O.}~\bibnamefont {Smirnova}}, \ and\ \bibinfo
		{author} {\bibfnamefont {V.}~\bibnamefont {Blanchet}},\ }\href {\doibase
		10.1038/s41567-017-0038-z} {\bibfield  {journal} {\bibinfo  {journal} {Nat.
				Phys.}\ }\textbf {\bibinfo {volume} {14}},\ \bibinfo {pages} {484} (\bibinfo
		{year} {2018})}\BibitemShut {NoStop}%
	\bibitem [{\citenamefont {Ranecky}\ \emph {et~al.}(2022)\citenamefont
		{Ranecky}, \citenamefont {Park}, \citenamefont {Samartzis}, \citenamefont
		{Giannakidis}, \citenamefont {Schwarzer}, \citenamefont {Senftleben},
		\citenamefont {Baumert},\ and\ \citenamefont {Schaefer}}]{Ranecky_PCCP_2022}%
	\BibitemOpen
	\bibfield  {author} {\bibinfo {author} {\bibfnamefont {S.~T.}\ \bibnamefont
			{Ranecky}}, \bibinfo {author} {\bibfnamefont {G.~B.}\ \bibnamefont {Park}},
		\bibinfo {author} {\bibfnamefont {P.~C.}\ \bibnamefont {Samartzis}}, \bibinfo
		{author} {\bibfnamefont {I.~C.}\ \bibnamefont {Giannakidis}}, \bibinfo
		{author} {\bibfnamefont {D.}~\bibnamefont {Schwarzer}}, \bibinfo {author}
		{\bibfnamefont {A.}~\bibnamefont {Senftleben}}, \bibinfo {author}
		{\bibfnamefont {T.}~\bibnamefont {Baumert}}, \ and\ \bibinfo {author}
		{\bibfnamefont {T.}~\bibnamefont {Schaefer}},\ }\href {\doibase
		10.1039/d1cp05468f} {\bibfield  {journal} {\bibinfo  {journal} {PCCP}\
		}\textbf {\bibinfo {volume} {24}},\ \bibinfo {pages} {2758} (\bibinfo {year}
		{2022})}\BibitemShut {NoStop}%
	\bibitem [{\citenamefont {Comby}\ \emph {et~al.}(2023)\citenamefont {Comby},
		\citenamefont {Descamps}, \citenamefont {Petit}, \citenamefont {Valzer},
		\citenamefont {Wloch}, \citenamefont {Pouysegu}, \citenamefont {Quideau},
		\citenamefont {Bockova}, \citenamefont {Meinert}, \citenamefont {Blanchet},
		\citenamefont {Fabre},\ and\ \citenamefont {Mairesse}}]{Comby_PCCP_2023}%
	\BibitemOpen
	\bibfield  {author} {\bibinfo {author} {\bibfnamefont {A.}~\bibnamefont
			{Comby}}, \bibinfo {author} {\bibfnamefont {D.}~\bibnamefont {Descamps}},
		\bibinfo {author} {\bibfnamefont {S.}~\bibnamefont {Petit}}, \bibinfo
		{author} {\bibfnamefont {E.}~\bibnamefont {Valzer}}, \bibinfo {author}
		{\bibfnamefont {M.}~\bibnamefont {Wloch}}, \bibinfo {author} {\bibfnamefont
			{L.}~\bibnamefont {Pouysegu}}, \bibinfo {author} {\bibfnamefont
			{S.}~\bibnamefont {Quideau}}, \bibinfo {author} {\bibfnamefont
			{J.}~\bibnamefont {Bockova}}, \bibinfo {author} {\bibfnamefont
			{C.}~\bibnamefont {Meinert}}, \bibinfo {author} {\bibfnamefont
			{V.}~\bibnamefont {Blanchet}}, \bibinfo {author} {\bibfnamefont
			{B.}~\bibnamefont {Fabre}}, \ and\ \bibinfo {author} {\bibfnamefont
			{Y.}~\bibnamefont {Mairesse}},\ }\href {\doibase 10.1039/d3cp01057k}
	{\bibfield  {journal} {\bibinfo  {journal} {PCCP}\ }\textbf {\bibinfo
			{volume} {25}},\ \bibinfo {pages} {16246} (\bibinfo {year}
		{2023})}\BibitemShut {NoStop}%
	\bibitem [{\citenamefont {Baykusheva}\ and\ \citenamefont
		{W\"orner}(2018)}]{BakyushevaPRX18}%
	\BibitemOpen
	\bibfield  {author} {\bibinfo {author} {\bibfnamefont {D.}~\bibnamefont
			{Baykusheva}}\ and\ \bibinfo {author} {\bibfnamefont {H.~J.}\ \bibnamefont
			{W\"orner}},\ }\href {\doibase 10.1103/PhysRevX.8.031060} {\bibfield
		{journal} {\bibinfo  {journal} {Phys. Rev. X}\ }\textbf {\bibinfo {volume}
			{8}},\ \bibinfo {pages} {031060} (\bibinfo {year} {2018})}\BibitemShut
	{NoStop}%
	\bibitem [{\citenamefont {Domingos}\ \emph {et~al.}(2020)\citenamefont
		{Domingos}, \citenamefont {Perez}, \citenamefont {Marshall}, \citenamefont
		{Leung},\ and\ \citenamefont {Schnell}}]{Domingos2020}%
	\BibitemOpen
	\bibfield  {author} {\bibinfo {author} {\bibfnamefont {S.~R.}\ \bibnamefont
			{Domingos}}, \bibinfo {author} {\bibfnamefont {C.}~\bibnamefont {Perez}},
		\bibinfo {author} {\bibfnamefont {M.~D.}\ \bibnamefont {Marshall}}, \bibinfo
		{author} {\bibfnamefont {H.~O.}\ \bibnamefont {Leung}}, \ and\ \bibinfo
		{author} {\bibfnamefont {M.}~\bibnamefont {Schnell}},\ }\href {\doibase
		{10.1039/d0sc03752d}} {\bibfield  {journal} {\bibinfo  {journal} {Chem.
				Sci.}\ }\textbf {\bibinfo {volume} {11}},\ \bibinfo {pages} {10863} (\bibinfo
		{year} {2020})}\BibitemShut {NoStop}%
	\bibitem [{\citenamefont {Owens}\ \emph {et~al.}(2018)\citenamefont {Owens},
		\citenamefont {Yachmenev}, \citenamefont {Yurchenko},\ and\ \citenamefont
		{K\"upper}}]{OwensPRL16}%
	\BibitemOpen
	\bibfield  {author} {\bibinfo {author} {\bibfnamefont {A.}~\bibnamefont
			{Owens}}, \bibinfo {author} {\bibfnamefont {A.}~\bibnamefont {Yachmenev}},
		\bibinfo {author} {\bibfnamefont {S.~N.}\ \bibnamefont {Yurchenko}}, \ and\
		\bibinfo {author} {\bibfnamefont {J.}~\bibnamefont {K\"upper}},\ }\href
	{\doibase 10.1103/PhysRevLett.121.193201} {\bibfield  {journal} {\bibinfo
			{journal} {Phys. Rev. Lett.}\ }\textbf {\bibinfo {volume} {121}},\ \bibinfo
		{pages} {193201} (\bibinfo {year} {2018})}\BibitemShut {NoStop}%
	\bibitem [{\citenamefont {Tikhonov}\ \emph {et~al.}(2022)\citenamefont
		{Tikhonov}, \citenamefont {Blech}, \citenamefont {Leibscher}, \citenamefont
		{Greeman}, \citenamefont {Schnell},\ and\ \citenamefont {Koch}}]{Tikhonov22}%
	\BibitemOpen
	\bibfield  {author} {\bibinfo {author} {\bibfnamefont {D.~S.}\ \bibnamefont
			{Tikhonov}}, \bibinfo {author} {\bibfnamefont {A.}~\bibnamefont {Blech}},
		\bibinfo {author} {\bibfnamefont {M.}~\bibnamefont {Leibscher}}, \bibinfo
		{author} {\bibfnamefont {L.}~\bibnamefont {Greeman}}, \bibinfo {author}
		{\bibfnamefont {M.}~\bibnamefont {Schnell}}, \ and\ \bibinfo {author}
		{\bibfnamefont {C.~P.}\ \bibnamefont {Koch}},\ }\href {\doibase
		10.1126/sciadv.ade0311} {\bibfield  {journal} {\bibinfo  {journal} {Sci.
				Adv.}\ }\textbf {\bibinfo {volume} {8}},\ \bibinfo {pages} {eade0311}
		(\bibinfo {year} {2022})}\BibitemShut {NoStop}%
	\bibitem [{\citenamefont {Ilchen}\ \emph {et~al.}(2017)\citenamefont {Ilchen},
		\citenamefont {Douguet}, \citenamefont {Mazza}, \citenamefont {Rafipoor},
		\citenamefont {Callegari}, \citenamefont {Finetti}, \citenamefont {Plekan},
		\citenamefont {Prince}, \citenamefont {Demidovich}, \citenamefont {Grazioli},
		\citenamefont {Avaldi}, \citenamefont {Bolognesi}, \citenamefont {Coreno},
		\citenamefont {Di~Fraia}, \citenamefont {Devetta}, \citenamefont
		{Ovcharenko}, \citenamefont {D\"usterer}, \citenamefont {Ueda}, \citenamefont
		{Bartschat}, \citenamefont {Grum-Grzhimailo}, \citenamefont {Bozhevolnov},
		\citenamefont {Kazansky}, \citenamefont {Kabachnik},\ and\ \citenamefont
		{Meyer}}]{Ilchen_PRL_2017}%
	\BibitemOpen
	\bibfield  {author} {\bibinfo {author} {\bibfnamefont {M.}~\bibnamefont
			{Ilchen}}, \bibinfo {author} {\bibfnamefont {N.}~\bibnamefont {Douguet}},
		\bibinfo {author} {\bibfnamefont {T.}~\bibnamefont {Mazza}}, \bibinfo
		{author} {\bibfnamefont {A.~J.}\ \bibnamefont {Rafipoor}}, \bibinfo {author}
		{\bibfnamefont {C.}~\bibnamefont {Callegari}}, \bibinfo {author}
		{\bibfnamefont {P.}~\bibnamefont {Finetti}}, \bibinfo {author} {\bibfnamefont
			{O.}~\bibnamefont {Plekan}}, \bibinfo {author} {\bibfnamefont {K.~C.}\
			\bibnamefont {Prince}}, \bibinfo {author} {\bibfnamefont {A.}~\bibnamefont
			{Demidovich}}, \bibinfo {author} {\bibfnamefont {C.}~\bibnamefont
			{Grazioli}}, \bibinfo {author} {\bibfnamefont {L.}~\bibnamefont {Avaldi}},
		\bibinfo {author} {\bibfnamefont {P.}~\bibnamefont {Bolognesi}}, \bibinfo
		{author} {\bibfnamefont {M.}~\bibnamefont {Coreno}}, \bibinfo {author}
		{\bibfnamefont {M.}~\bibnamefont {Di~Fraia}}, \bibinfo {author}
		{\bibfnamefont {M.}~\bibnamefont {Devetta}}, \bibinfo {author} {\bibfnamefont
			{Y.}~\bibnamefont {Ovcharenko}}, \bibinfo {author} {\bibfnamefont
			{S.}~\bibnamefont {D\"usterer}}, \bibinfo {author} {\bibfnamefont
			{K.}~\bibnamefont {Ueda}}, \bibinfo {author} {\bibfnamefont {K.}~\bibnamefont
			{Bartschat}}, \bibinfo {author} {\bibfnamefont {A.~N.}\ \bibnamefont
			{Grum-Grzhimailo}}, \bibinfo {author} {\bibfnamefont {A.~V.}\ \bibnamefont
			{Bozhevolnov}}, \bibinfo {author} {\bibfnamefont {A.~K.}\ \bibnamefont
			{Kazansky}}, \bibinfo {author} {\bibfnamefont {N.~M.}\ \bibnamefont
			{Kabachnik}}, \ and\ \bibinfo {author} {\bibfnamefont {M.}~\bibnamefont
			{Meyer}},\ }\href {\doibase 10.1103/PhysRevLett.118.013002} {\bibfield
		{journal} {\bibinfo  {journal} {Phys. Rev. Lett.}\ }\textbf {\bibinfo
			{volume} {118}},\ \bibinfo {pages} {013002} (\bibinfo {year}
		{2017})}\BibitemShut {NoStop}%
	\bibitem [{\citenamefont {Ordonez}\ and\ \citenamefont
		{Smirnova}(2019{\natexlab{b}})}]{Ordonez_PRA_2019_1}%
	\BibitemOpen
	\bibfield  {author} {\bibinfo {author} {\bibfnamefont {A.~F.}\ \bibnamefont
			{Ordonez}}\ and\ \bibinfo {author} {\bibfnamefont {O.}~\bibnamefont
			{Smirnova}},\ }\href {\doibase 10.1103/PhysRevA.99.043416} {\bibfield
		{journal} {\bibinfo  {journal} {Phys. Rev. A}\ }\textbf {\bibinfo {volume}
			{99}},\ \bibinfo {pages} {043416} (\bibinfo {year}
		{2019}{\natexlab{b}})}\BibitemShut {NoStop}%
	\bibitem [{\citenamefont {Grum-Grzhimailo}\ \emph {et~al.}(2019)\citenamefont
		{Grum-Grzhimailo}, \citenamefont {Douguet}, \citenamefont {Meyer},\ and\
		\citenamefont {Bartschat}}]{GrumG_PRA_2019}%
	\BibitemOpen
	\bibfield  {author} {\bibinfo {author} {\bibfnamefont {A.~N.}\ \bibnamefont
			{Grum-Grzhimailo}}, \bibinfo {author} {\bibfnamefont {N.}~\bibnamefont
			{Douguet}}, \bibinfo {author} {\bibfnamefont {M.}~\bibnamefont {Meyer}}, \
		and\ \bibinfo {author} {\bibfnamefont {K.}~\bibnamefont {Bartschat}},\ }\href
	{\doibase 10.1103/PhysRevA.100.033404} {\bibfield  {journal} {\bibinfo
			{journal} {Phys. Rev. A}\ }\textbf {\bibinfo {volume} {100}},\ \bibinfo
		{pages} {033404} (\bibinfo {year} {2019})}\BibitemShut {NoStop}%
	\bibitem [{\citenamefont {Buhmann}\ \emph {et~al.}(2021)\citenamefont
		{Buhmann}, \citenamefont {Giesen}, \citenamefont {Diekmann}, \citenamefont
		{Berger}, \citenamefont {Aull}, \citenamefont {Zahariev}, \citenamefont
		{Debatin},\ and\ \citenamefont {Singer}}]{Buhmann_2021}%
	\BibitemOpen
	\bibfield  {author} {\bibinfo {author} {\bibfnamefont {S.~Y.}\ \bibnamefont
			{Buhmann}}, \bibinfo {author} {\bibfnamefont {S.~M.}\ \bibnamefont {Giesen}},
		\bibinfo {author} {\bibfnamefont {M.}~\bibnamefont {Diekmann}}, \bibinfo
		{author} {\bibfnamefont {R.}~\bibnamefont {Berger}}, \bibinfo {author}
		{\bibfnamefont {S.}~\bibnamefont {Aull}}, \bibinfo {author} {\bibfnamefont
			{P.}~\bibnamefont {Zahariev}}, \bibinfo {author} {\bibfnamefont
			{M.}~\bibnamefont {Debatin}}, \ and\ \bibinfo {author} {\bibfnamefont
			{K.}~\bibnamefont {Singer}},\ }\href {\doibase 10.1088/1367-2630/ac1af7}
	{\bibfield  {journal} {\bibinfo  {journal} {New Journal of Physics}\ }\textbf
		{\bibinfo {volume} {23}},\ \bibinfo {pages} {083040} (\bibinfo {year}
		{2021})}\BibitemShut {NoStop}%
	\bibitem [{\citenamefont {Mayer}\ \emph {et~al.}(2022)\citenamefont {Mayer},
		\citenamefont {Patchkovskii}, \citenamefont {Morales}, \citenamefont
		{Ivanov},\ and\ \citenamefont {Smirnova}}]{MayerPRL22}%
	\BibitemOpen
	\bibfield  {author} {\bibinfo {author} {\bibfnamefont {N.}~\bibnamefont
			{Mayer}}, \bibinfo {author} {\bibfnamefont {S.}~\bibnamefont {Patchkovskii}},
		\bibinfo {author} {\bibfnamefont {F.}~\bibnamefont {Morales}}, \bibinfo
		{author} {\bibfnamefont {M.}~\bibnamefont {Ivanov}}, \ and\ \bibinfo {author}
		{\bibfnamefont {O.}~\bibnamefont {Smirnova}},\ }\href {\doibase
		10.1103/PhysRevLett.129.243201} {\bibfield  {journal} {\bibinfo  {journal}
			{Phys. Rev. Lett.}\ }\textbf {\bibinfo {volume} {129}},\ \bibinfo {pages}
		{243201} (\bibinfo {year} {2022})}\BibitemShut {NoStop}%
	\bibitem [{\citenamefont {Ordonez}\ and\ \citenamefont
		{Smirnova}(2018)}]{OrdonezPRA18}%
	\BibitemOpen
	\bibfield  {author} {\bibinfo {author} {\bibfnamefont {A.~F.}\ \bibnamefont
			{Ordonez}}\ and\ \bibinfo {author} {\bibfnamefont {O.}~\bibnamefont
			{Smirnova}},\ }\href {\doibase 10.1103/PhysRevA.98.063428} {\bibfield
		{journal} {\bibinfo  {journal} {Phys. Rev. A}\ }\textbf {\bibinfo {volume}
			{98}},\ \bibinfo {pages} {063428} (\bibinfo {year} {2018})}\BibitemShut
	{NoStop}%
	\bibitem [{\citenamefont {Bunker}\ and\ \citenamefont {Jensen}(1998)}]{Bunker}%
	\BibitemOpen
	\bibfield  {author} {\bibinfo {author} {\bibfnamefont {P.~R.}\ \bibnamefont
			{Bunker}}\ and\ \bibinfo {author} {\bibfnamefont {P.}~\bibnamefont
			{Jensen}},\ }\href@noop {} {\emph {\bibinfo {title} {Molecular Symmetry and
				Spectroscopy}}}\ (\bibinfo  {publisher} {NRC Research Press},\ \bibinfo
	{year} {1998})\BibitemShut {NoStop}%
	\bibitem [{\citenamefont {D'Alessandro}(2008)}]{Alessandro2008}%
	\BibitemOpen
	\bibfield  {author} {\bibinfo {author} {\bibfnamefont {D.}~\bibnamefont
			{D'Alessandro}},\ }\href@noop {} {\emph {\bibinfo {title} {Quantum Control
				and Dynamics}}}\ (\bibinfo  {publisher} {Chapman and Hall},\ \bibinfo {year}
	{2008})\BibitemShut {NoStop}%
	\bibitem [{\citenamefont {Judson}\ \emph {et~al.}(1990)\citenamefont {Judson},
		\citenamefont {Lehmann}, \citenamefont {Rabitz},\ and\ \citenamefont
		{Warren}}]{Judson1990}%
	\BibitemOpen
	\bibfield  {author} {\bibinfo {author} {\bibfnamefont {R.}~\bibnamefont
			{Judson}}, \bibinfo {author} {\bibfnamefont {K.}~\bibnamefont {Lehmann}},
		\bibinfo {author} {\bibfnamefont {H.}~\bibnamefont {Rabitz}}, \ and\ \bibinfo
		{author} {\bibfnamefont {W.}~\bibnamefont {Warren}},\ }\href@noop {}
	{\bibfield  {journal} {\bibinfo  {journal} {J. Mol. Struct.}\ }\textbf
		{\bibinfo {volume} {223}},\ \bibinfo {pages} {425} (\bibinfo {year}
		{1990})}\BibitemShut {NoStop}%
	\bibitem [{\citenamefont {Boscain}\ \emph {et~al.}(2014)\citenamefont
		{Boscain}, \citenamefont {Caponigro},\ and\ \citenamefont {Sigalotti}}]{BCS}%
	\BibitemOpen
	\bibfield  {author} {\bibinfo {author} {\bibfnamefont {U.}~\bibnamefont
			{Boscain}}, \bibinfo {author} {\bibfnamefont {M.}~\bibnamefont {Caponigro}},
		\ and\ \bibinfo {author} {\bibfnamefont {M.}~\bibnamefont {Sigalotti}},\
	}\href {\doibase 10.1016/j.jde.2014.02.004} {\bibfield  {journal} {\bibinfo
			{journal} {J. Differ. Equ.}\ }\textbf {\bibinfo {volume} {256}},\ \bibinfo
		{pages} {3524} (\bibinfo {year} {2014})}\BibitemShut {NoStop}%
	\bibitem [{\citenamefont {Chambrion}\ and\ \citenamefont
		{Pozzoli}(2023)}]{Chambrion23}%
	\BibitemOpen
	\bibfield  {author} {\bibinfo {author} {\bibfnamefont {T.}~\bibnamefont
			{Chambrion}}\ and\ \bibinfo {author} {\bibfnamefont {E.}~\bibnamefont
			{Pozzoli}},\ }\href {\doibase
		https://doi.org/10.1016/j.automatica.2023.111028} {\bibfield  {journal}
		{\bibinfo  {journal} {Automatica}\ }\textbf {\bibinfo {volume} {153}},\
		\bibinfo {pages} {111028} (\bibinfo {year} {2023})}\BibitemShut {NoStop}%
	\bibitem [{\citenamefont {Boscain}\ \emph {et~al.}(2021)\citenamefont
		{Boscain}, \citenamefont {Pozzoli},\ and\ \citenamefont
		{Sigalotti}}]{Boscain21}%
	\BibitemOpen
	\bibfield  {author} {\bibinfo {author} {\bibfnamefont {U.}~\bibnamefont
			{Boscain}}, \bibinfo {author} {\bibfnamefont {E.}~\bibnamefont {Pozzoli}}, \
		and\ \bibinfo {author} {\bibfnamefont {M.}~\bibnamefont {Sigalotti}},\ }\href
	{\doibase 10.1137/20M1311442} {\bibfield  {journal} {\bibinfo  {journal}
			{SIAM J. Control Optim.}\ }\textbf {\bibinfo {volume} {59}},\ \bibinfo
		{pages} {156} (\bibinfo {year} {2021})}\BibitemShut {NoStop}%
	\bibitem [{\citenamefont {Chambrion}\ and\ \citenamefont
		{Pozzoli}(2022)}]{Chambrion22}%
	\BibitemOpen
	\bibfield  {author} {\bibinfo {author} {\bibfnamefont {T.}~\bibnamefont
			{Chambrion}}\ and\ \bibinfo {author} {\bibfnamefont {E.}~\bibnamefont
			{Pozzoli}},\ }\href {\doibase 10.1109/LCSYS.2022.3162250} {\bibfield
		{journal} {\bibinfo  {journal} {IEEE Control Systems Letters}\ }\textbf
		{\bibinfo {volume} {6}},\ \bibinfo {pages} {2425} (\bibinfo {year}
		{2022})}\BibitemShut {NoStop}%
	\bibitem [{\citenamefont {Leibscher}\ \emph {et~al.}(2022)\citenamefont
		{Leibscher}, \citenamefont {Pozzoli}, \citenamefont {Perez}, \citenamefont
		{Schnell}, \citenamefont {Sigalotti}, \citenamefont {Boscain},\ and\
		\citenamefont {Koch}}]{Leibscher22}%
	\BibitemOpen
	\bibfield  {author} {\bibinfo {author} {\bibfnamefont {M.}~\bibnamefont
			{Leibscher}}, \bibinfo {author} {\bibfnamefont {E.}~\bibnamefont {Pozzoli}},
		\bibinfo {author} {\bibfnamefont {C.}~\bibnamefont {Perez}}, \bibinfo
		{author} {\bibfnamefont {M.}~\bibnamefont {Schnell}}, \bibinfo {author}
		{\bibfnamefont {M.}~\bibnamefont {Sigalotti}}, \bibinfo {author}
		{\bibfnamefont {U.}~\bibnamefont {Boscain}}, \ and\ \bibinfo {author}
		{\bibfnamefont {C.~P.}\ \bibnamefont {Koch}},\ }\href
	{https://doi.org/10.1038/s42005-022-00883-6} {\bibfield  {journal} {\bibinfo
			{journal} {Commun. Phys.}\ }\textbf {\bibinfo {volume} {5}} (\bibinfo {year}
		{2022})}\BibitemShut {NoStop}%
	\bibitem [{\citenamefont {Pozzoli}\ \emph {et~al.}(2022)\citenamefont
		{Pozzoli}, \citenamefont {Leibscher}, \citenamefont {Sigalotti},
		\citenamefont {Boscain},\ and\ \citenamefont {Koch}}]{Pozzoli21}%
	\BibitemOpen
	\bibfield  {author} {\bibinfo {author} {\bibfnamefont {E.}~\bibnamefont
			{Pozzoli}}, \bibinfo {author} {\bibfnamefont {M.}~\bibnamefont {Leibscher}},
		\bibinfo {author} {\bibfnamefont {M.}~\bibnamefont {Sigalotti}}, \bibinfo
		{author} {\bibfnamefont {U.}~\bibnamefont {Boscain}}, \ and\ \bibinfo
		{author} {\bibfnamefont {C.~P.}\ \bibnamefont {Koch}},\ }\href {\doibase
		10.1088/1751-8121/ac631d} {\bibfield  {journal} {\bibinfo  {journal} {J.
				Phys. A: Math. Theo.}\ }\textbf {\bibinfo {volume} {55}} (\bibinfo {year}
		{2022}),\ 10.1088/1751-8121/ac631d}\BibitemShut {NoStop}%
	\bibitem [{\citenamefont {Pozzoli}(2022)}]{Pozzoli22}%
	\BibitemOpen
	\bibfield  {author} {\bibinfo {author} {\bibfnamefont {E.}~\bibnamefont
			{Pozzoli}},\ }\href {\doibase 10.1007/s00245-022-09821-y} {\bibfield
		{journal} {\bibinfo  {journal} {Appl. Math. Optim.}\ }\textbf {\bibinfo
			{volume} {85}},\ \bibinfo {pages} {Paper No. 8, 27} (\bibinfo {year}
		{2022})}\BibitemShut {NoStop}%
	\bibitem [{\citenamefont {Boscain}\ \emph {et~al.}(2012)\citenamefont
		{Boscain}, \citenamefont {Caponigro}, \citenamefont {Chambrion},\ and\
		\citenamefont {Sigalotti}}]{BCCS}%
	\BibitemOpen
	\bibfield  {author} {\bibinfo {author} {\bibfnamefont {U.}~\bibnamefont
			{Boscain}}, \bibinfo {author} {\bibfnamefont {M.}~\bibnamefont {Caponigro}},
		\bibinfo {author} {\bibfnamefont {T.}~\bibnamefont {Chambrion}}, \ and\
		\bibinfo {author} {\bibfnamefont {M.}~\bibnamefont {Sigalotti}},\ }\href
	{\doibase 10.1007/s00220-012-1441-z} {\bibfield  {journal} {\bibinfo
			{journal} {Comm. Math. Phys.}\ }\textbf {\bibinfo {volume} {311}},\ \bibinfo
		{pages} {423} (\bibinfo {year} {2012})}\BibitemShut {NoStop}%
	\bibitem [{\citenamefont {Chambrion}\ \emph {et~al.}(2009)\citenamefont
		{Chambrion}, \citenamefont {Mason}, \citenamefont {Sigalotti},\ and\
		\citenamefont {Boscain}}]{CMSB}%
	\BibitemOpen
	\bibfield  {author} {\bibinfo {author} {\bibfnamefont {T.}~\bibnamefont
			{Chambrion}}, \bibinfo {author} {\bibfnamefont {P.}~\bibnamefont {Mason}},
		\bibinfo {author} {\bibfnamefont {M.}~\bibnamefont {Sigalotti}}, \ and\
		\bibinfo {author} {\bibfnamefont {U.}~\bibnamefont {Boscain}},\ }\href
	{\doibase 10.1016/j.anihpc.2008.05.001} {\bibfield  {journal} {\bibinfo
			{journal} {Ann. Inst. H. Poincar\'{e} Anal. Non Lin\'{e}aire}\ }\textbf
		{\bibinfo {volume} {26}},\ \bibinfo {pages} {329} (\bibinfo {year}
		{2009})}\BibitemShut {NoStop}%
	\bibitem [{\citenamefont {Wang}\ \emph {et~al.}(2023)\citenamefont {Wang},
		\citenamefont {Li}, \citenamefont {Li}, \citenamefont {Petersen},\ and\
		\citenamefont {Shi}}]{graphs}%
	\BibitemOpen
	\bibfield  {author} {\bibinfo {author} {\bibfnamefont {X.}~\bibnamefont
			{Wang}}, \bibinfo {author} {\bibfnamefont {B.}~\bibnamefont {Li}}, \bibinfo
		{author} {\bibfnamefont {J.-S.}\ \bibnamefont {Li}}, \bibinfo {author}
		{\bibfnamefont {I.~R.}\ \bibnamefont {Petersen}}, \ and\ \bibinfo {author}
		{\bibfnamefont {G.}~\bibnamefont {Shi}},\ }\href@noop {} {\bibfield
		{journal} {\bibinfo  {journal} {IEEE Trans. Automat. Control}\ }\textbf
		{\bibinfo {volume} {68}},\ \bibinfo {pages} {2277} (\bibinfo {year}
		{2023})}\BibitemShut {NoStop}%
	\bibitem [{Note1()}]{Note1}%
	\BibitemOpen
	\bibinfo {note} {Each asymmetric top eigenfunction is uniquely described by
		$J$, $M$ and the two corresponding symmetric top quantum numbers $K_a$ and
		$K_c$. Since the rotational energy eigenvalues $E_j^{rot}
		=E_{J_{K_a,K_c}}^{rot}$ do not depend on $M$, we denote the rotational energy
		levels by ${J_{K_a,K_c}}$.}\BibitemShut {Stop}%
	\bibitem [{\citenamefont {Gago-Encinas}\ \emph {et~al.}(2023)\citenamefont
		{Gago-Encinas}, \citenamefont {Leibscher},\ and\ \citenamefont
		{Koch}}]{Gago_2023}%
	\BibitemOpen
	\bibfield  {author} {\bibinfo {author} {\bibfnamefont {F.}~\bibnamefont
			{Gago-Encinas}}, \bibinfo {author} {\bibfnamefont {M.}~\bibnamefont
			{Leibscher}}, \ and\ \bibinfo {author} {\bibfnamefont {C.~P.}\ \bibnamefont
			{Koch}},\ }\href {\doibase 10.1088/2058-9565/ace1a4} {\bibfield  {journal}
		{\bibinfo  {journal} {Quantum Sci. Technol.}\ }\textbf {\bibinfo {volume}
			{8}},\ \bibinfo {pages} {045002} (\bibinfo {year} {2023})}\BibitemShut
	{NoStop}%
	\bibitem [{Note2()}]{Note2}%
	\BibitemOpen
	\bibinfo {note} {{\protect \color {black} Actually, this condition implies a
			stronger notion of controllability, namely, controllability at the level of
			the propagators, which implies, in particular, controllability of the density
			matrices.}}\BibitemShut {Stop}%
	\bibitem [{\citenamefont {Zare}(1988)}]{Zare88}%
	\BibitemOpen
	\bibfield  {author} {\bibinfo {author} {\bibfnamefont {R.~N.}\ \bibnamefont
			{Zare}},\ }\href@noop {} {\emph {\bibinfo {title} {Angular Momentum}}}\
	(\bibinfo  {publisher} {Wiley},\ \bibinfo {year} {1988})\BibitemShut
	{NoStop}%
\end{thebibliography}
%

\end{document}